\documentclass[review, 3p, 8pt, times, authoryear,fleqn]{elsarticle}
\usepackage{amsmath,amssymb,xspace,wrapfig,bm}
\usepackage{svg}  
\usepackage{tabu} 
\usepackage{array} 
\usepackage{float}
\usepackage{subfigure}
\usepackage{xcolor}  
\usepackage[colorlinks=true,citecolor=blue]{hyperref}
\usepackage[nameinlink,capitalise]{cleveref} 
\usepackage{appendix}
\usepackage{lineno} 
\usepackage{dirtytalk} 
\usepackage{ulem} 
\usepackage{booktabs} 
\journal{Composites Part B: Engineering}  
\makeatletter
\def\ps@pprintTitle{%
  \let\@oddhead\@empty
  \let\@evenhead\@empty
  \let\@oddfoot\@empty
  \let\@evenfoot\@oddfoot
}
\makeatother 
\newcommand{\argminWithArgs}{\operatornamewithlimits{arg\ min}}

\newcommand{\JB}[1]{\textcolor{cyan}{#1}}

\begin{document}
\begin{frontmatter}
\title{Single-test evaluation of directional elastic properties of anisotropic structured materials}

\author[1]{Jagannadh Boddapati\fnref{ec}}
\author[2]{Moritz Flaschel\fnref{ec}}
\author[3]{Siddhant Kumar}
\author[2]{Laura De Lorenzis}
\author[1]{Chiara Daraio\corref{cor1}}
\ead{daraio@caltech.edu}

\cortext[cor1]{Corresponding author}
\fntext[ec]{These authors contributed equally.}

\address[1]{Division of Engineering and Applied Science, California Institute of Technology, Pasadena, CA 91125, USA}
\address[2]{Department of Mechanical and Process Engineering, ETH Zürich, 8092 Zürich, Switzerland}
\address[3]{Department of Materials Science and Engineering, Delft University of Technology, 2628 CD Delft, The Netherlands}

\begin{abstract}
When the elastic properties of structured materials become direction-dependent, the number of their descriptors increases.
For example, in two-dimensions, the anisotropic behavior of materials is described by up to 6 independent elastic stiffness parameters, as opposed to only 2 needed for isotropic materials. 
Such high number of parameters expands the design space of structured materials and leads to unusual phenomena, such as materials that can shear under uniaxial compression. 
However, an increased number of properties descriptors and the coupling between shear and normal deformations render the experimental evaluation of material properties more challenging.
In this paper, we propose a methodology based on the virtual fields method to identify six separate stiffness tensor parameters of two-dimensional anisotropic structured materials using just one tension test, thus eliminating the need for multiple experiments, as it is typical in traditional methods. The approach requires no stress data and uses full-field displacement data and global force data.
{We show the accuracy of our method using synthetic data generated from finite element simulations as well as experimental data from additively manufactured specimens.}
\end{abstract}


\end{frontmatter}

\newpage
\section{Introduction}
The advent of additive manufacturing has allowed the design and engineering of a new class of materials known as metamaterials, or structured/architected materials.
 Mechanical metamaterials are a special branch of metamaterials that derive special functionalities from their peculiar deformation, dynamic motion and/or elastic energy distribution \citep{lee_micro-nanostructured_2012, christensen_vibrant_2015,zadpoor_mechanical_2016,bertoldi_flexible_2017, surjadi_mechanical_2019}.
 Metamaterials derive their effective properties from both the micro- and meso-structure and their constitutive material properties. They often exhibit mechanical properties that deviate from those of their constituent materials, showing unusual behaviors, such as negative Poisson's ratios \citep{greaves_poissons_2011}, vanishing shear moduli \citep{kadic_practicability_2012}, and negative refractive indices \citep{kaina_negative_2015}.

By carefully selecting the geometry of the micro- and meso-structures with varying symmetries \citep{milton_which_1995,kadic_practicability_2012,wu_mechanical_2019, kulagin_architectured_2020,mao_twist_2020,bastek_inverting_2022}, metamaterial designers can explore novel anisotropy classes in the material responses.
In turn, the presence of rich anisotropy expands the materials' functionality space, by exploiting coupled-deformation mechanisms that are non-existent in symmetric structures. 
Examples include metamaterials that twist under compression \citep{frenzel_three-dimensional_2017,chen_optimization_2018,wu_mechanical_2019,yuan_micropolar_2021}, shear under thermal loading \citep{ni_2d_2019} and shape-morph \citep{guseinov_programming_2020,risso_instability-driven_2021,agnelli_design_2022}. 
In the dynamic regime, anisotropy allows observing phenomena like conical refraction \citep{ahn_conical_2017} and control of broadband elastic waves \citep{zheng_theory_2019,yang_monolayer_2019,zheng_non-resonant_2020}. 

In a two-dimensional continuum, the elastic behavior of an anisotropic material is described using six independent elastic parameters \citep{ting1996anisotropic}. 
In experiments, characterizing these many independent elastic parameters is quite complex. 
Indeed, the presence of shear-normal coupling makes it hard to measure even one of the six parameters from a single experiment.
Prior work suggested different approaches to experimentally measure the elastic parameters for different anisotropy classes \citep{schittny_elastic_2013,gras_identification_2015,lee_effective_2016,kim_virtual_2020,agnelli_systematic_2021}.
However, most of these approaches focus on measuring the stiffness tensor components when the off-diagonal, shear-normal coupling, components are absent.
In addition, several of these approaches require multiple experimental steps. For example, techniques based on the detection of different acoustic wave speeds along different material directions involve multiple tests and assume a certain material symmetry in predicting elastic parameters \citep{every_determination_1990, francois_determination_1998}. 
To date, there are no experimental methods that can measure the stiffness parameters of fully anisotropic structures from a single experiment. 

Traditional material parameter identification methods rely on single-load experimental setups with homogeneous (constant) strain distributions within the tested specimen, which allow the derivation of closed-form stress-strain relations.
However, the amount of data that can be acquired through a one dimensional tension test, for example, is limited {(}e.g., one stress-strain data pair for each measurement{)}. 
When characterizing complex materials, multiple experimental setups with different loading conditions are needed.
Full-field identification methods allow extracting additional information from single-load experiments. Measuring the full displacement field, e.g., through Digital Image Correlation (DIC), of arbitrarily shaped specimens under loading maximizes the amount of data generated from a single experimental test.
Such data can then be used to characterize the material by applying inverse identification methods such as, among others, Finite Element Model Updating, the Equilibrium Gap Method or the Virtual Fields Method (VFM), see \citep{avril_overview_2008,roux_optimal_2020,pierron_material_nodate} for a review.

These methods have in common that they are used to calibrate the parameters of an a priori chosen material model, i.e., the mathematical functions and operations that describe the material response need to be fixed by means of the intuition or modeling experience of the user. However, the selection of inappropriate a priori assumptions about the model and its underlying mathematical structure can introduce errors. Recent research used full-field data to train machine-learning-models, whose versatile ansatz spaces promise to mitigate modeling errors.
\cite{flaschel_unsupervised_2021}, for example, proposed the method EUCLID (Efficient Unsupervised Constitutive Law Identification and Discovery) that uses sparse regression \citep{tibshirani_regression_1996} informed by full-field displacement data and net reaction force data, to automatically select interpretable material models from a potentially large predefined set of candidate material models.
EUCLID has been applied to hyperelasticity \citep{flaschel_unsupervised_2021}, elastoplasticity \citep{flaschel_discovering_2022}, viscoelasticity \citep{marino_automated_2023}, and generalized standard materials \citep{flaschel_automated_2023}, see \cite{flaschel_automated_2023-1} for an overview.
Further, EUCLID was formulated in a Bayesian setting by \cite{joshi_bayesian-euclid_2022} to simultaneously perform model selection and quantification of uncertainty in the material parameters.
In contrast to selecting interpretable material models through sparse regression, full-field data may also be used to train black-box material model surrogates like neural networks, as shown by \cite{man_neural_2011,huang_learning_2020,liu_learning_2020} for small strain elasticity and by \cite{thakolkaran_nn-euclid_2022} for hyperelasticity.
In the present work, it is assumed that the material response does not leave the realm of elasticity at infinitesimal strains.
Thus, the material model can be assumed to be known a priori, and its parameters are calibrated with the VFM.

The VFM, originally proposed by \cite{grediac_principle_1989} (see also \cite{grediac_virtual_2008,pierron_virtual_2012}), employs the balance of linear momentum in its weak form, to identify unknown material parameters. 
The VFM method assumes that the kinematic fields in the specimen, as well as the reaction forces at the boundaries, are known from experiments.
As such, material parameters remain the only unknowns in the balance equations and can be calculated using standard linear or nonlinear solvers.
In essence, the VFM describes the inverse problem to the classical Finite Element Method (FEM). 
The method has been applied in various cases, such as small-strain elasticity, elasto-plasticity \citep{grediac_applying_2006}, and hyperelasticity \citep{promma_application_2009}, among others.

The accuracy of the VFM in identifying unknown material parameters and its sensitivity to noise are highly dependent on the choice of the functions for which the weak linear momentum balance is tested, also known as the virtual displacement fields. 
A distinction can be made between \textit{global} virtual fields that are defined over the whole specimen domain, such as polynomials, and \textit{local} virtual fields with compact support, such as in the Bubnov-Galerkin discretization with piecewise polynomial shape functions.
As the choice of the virtual fields is arbitrary and user-dependent, several attempts have been made to automate and optimize it \citep{avril_sensitivity_2004,pierron_extension_2010,marek_sensitivity-based_2017}.

In this article, full-field measurement based identification, and in particular the VFM, is explored in the context of anisotropic structured materials and compared to traditional identification methods. 
We focus in particular on the identification of shear-normal coupling parameters, notoriously complex to extract from conventional experiments.
The rest of the paper is organized as follows. In \cref{sec: theory}, we discuss the theory of anisotropic linear elasticity and introduce our model setup used for parameter identification.
In \cref{sec: VFM}, we present our virtual fields method.
In \cref{sec: data aq}, we describe our experimental and numerical data acquisition methods.
In \cref{sec: rd}, we discuss our results, including experimental validation, and we draw our conclusions in \cref{sec: concl}.

\section{Material model and geometry}
\label{sec: theory}
In this section, we review the fundamental equations of linear elasticity at infinitesimal strains and introduce our model setup used to identify the governing material parameters of anisotropic metamaterials.

\subsection{Anisotropic linear elasticity}
Under the small strain assumption, the constitutive law for a general anisotropic solid, which relates the Cauchy stress tensor $\bm{\sigma}$ and the infinitesimal strain tensor $\bm{\varepsilon}$,  is given by the generalized Hooke's law \citep{rychlewski_hookes_1984,ting1996anisotropic},
\begin{equation}
\bm{\sigma} =  \bm{C\varepsilon} \hspace{10 pt }\text{or} \hspace{10 pt } (\sigma_{ij} = C_{ijkl}\varepsilon_{kl}), \label{eq: 3D Hooke's law}
\end{equation} 
where \bm{$C$} is a fourth-order tensor, known as the elasticity tensor or the stiffness tensor, and Einstein's notation for summation over repeated indices is followed. 
For a two-dimensional anisotropic solid, under plane stress conditions, \cref{eq: 3D Hooke's law} can be written using Voigt notation as
\begin{equation}
\left[\begin{array}{l}
\sigma_{11} \\
\sigma_{22} \\
\sigma_{12}
\end{array}\right]=\left[\begin{array}{lll}
C_{1111} & C_{1122} & C_{1112} \\
C_{1122} & C_{2222} & C_{2212} \\
C_{1112} & C_{2212} & C_{1212} \\
\end{array}\right]\left[\begin{array}{c}
\varepsilon_{11} \\
\varepsilon_{22} \\
2 \varepsilon_{12}
\end{array}\right], \label{eq: ConstIndicial}
\end{equation}
where $C_{1111}, \ C_{1122}, \ C_{2222}, \ C_{1112}, \ C_{2212}, \ C_{1212}$ are the elasticity tensor parameters in a given reference frame,
$\varepsilon_{11}, \varepsilon_{22}$ are the axial strains, $\varepsilon_{12}$ is the shear strain,  $\sigma_{11}, \sigma_{22}$ are the axial stresses, and $\sigma_{12}$ is the shear stress.
For readability, we combine the pair of indices as follows: $()_{11} \rightarrow()_1, ()_{22} \rightarrow()_2, ()_{12} \rightarrow()_6$ and write \cref{eq: ConstIndicial} as
\begin{equation}
\left[\begin{array}{l}
\sigma_{{1}} \\
\sigma_{{2}} \\
\sigma_{{6}}
\end{array}\right]=\left[\begin{array}{lll}
C_{11} & C_{12} & C_{16} \\
C_{12} & C_{22} & C_{26} \\
C_{16} & C_{26} & C_{66} \\
\end{array}\right]\left[\begin{array}{c}
\varepsilon_{{1}} \\
\varepsilon_{{2}} \\
2 \varepsilon_{{6}}
\end{array}\right]. \label{eq: ConstIndicialSmall}
\end{equation}

Our objective is to identify these six material parameters $C_{11}, C_{12}, C_{22}, C_{16}, C_{26}, C_{66}$ from experimental measurements while fulfilling certain constraints.
From thermodynamic constraints, the elasticity tensor has to be positive definite, which implies
\begin{subequations}
\begin{align}
&C_{11} > 0, \quad C_{22}> 0, \quad C_{66} > 0, \\
&C_{11}C_{22}-C_{12}^2 >0, \quad C_{11}C_{66}-C_{16}^2 >0, \quad C_{22}C_{66}-C_{26}^2 >0.
\end{align}
\end{subequations}

The stiffness parameter $C_{12}$ represents the extension-to-extension deformation coupling. 
The stiffness parameters $C_{16},C_{26}$ represent the extension-to-shear coupling, also known as shear-normal coupling, which induces shear stress from axial strains, and axial stresses from shear strains.
Shear-normal coupling has been explored in the context of structured materials by \cite{karathanasopoulos_mechanics_2020, dos_reis_inverse_2022}. 
As a result of these anisotropy-induced couplings, the experimental identification of the material parameters becomes non-trivial because a constant state of strain is hard to achieve, even in a standard uniaxial tension test.

Note that the parameters $C_{16}$ and $C_{26}$ will be zero if the material has symmetry planes along the $x_1$ and $x_2$ axes.
Thus, the existence of shear-normal coupling and the maximum number of independent stiffness tensor parameters depend on the symmetries associated with the material microscopic topology \citep{ting1996anisotropic, podesta_symmetry_2019}. 
In plane elasticity, stiffness tensors are categorized into four symmetry classes. 
They are denoted as $O(2)$ for Isotropic, $D_4$ for Tetragonal, $D_2$ for Orthotropic and $Z_2$ for Digonal (fully anisotropic) with 2, 3, 4 and 6 independent parameters respectively. 
This categorization is based on the invariants of the stiffness tensor \citep{forte_unified_2014,auffray_invariant-based_2016}. 
However, in our methods of parameter identification, we do not consider any prior information on the material symmetries or the number of independent material parameters.

\subsection{Model setup}
\begin{figure}[!htb]
	\begin{center}
		\centering 
		\includegraphics[width =0.8\textwidth]{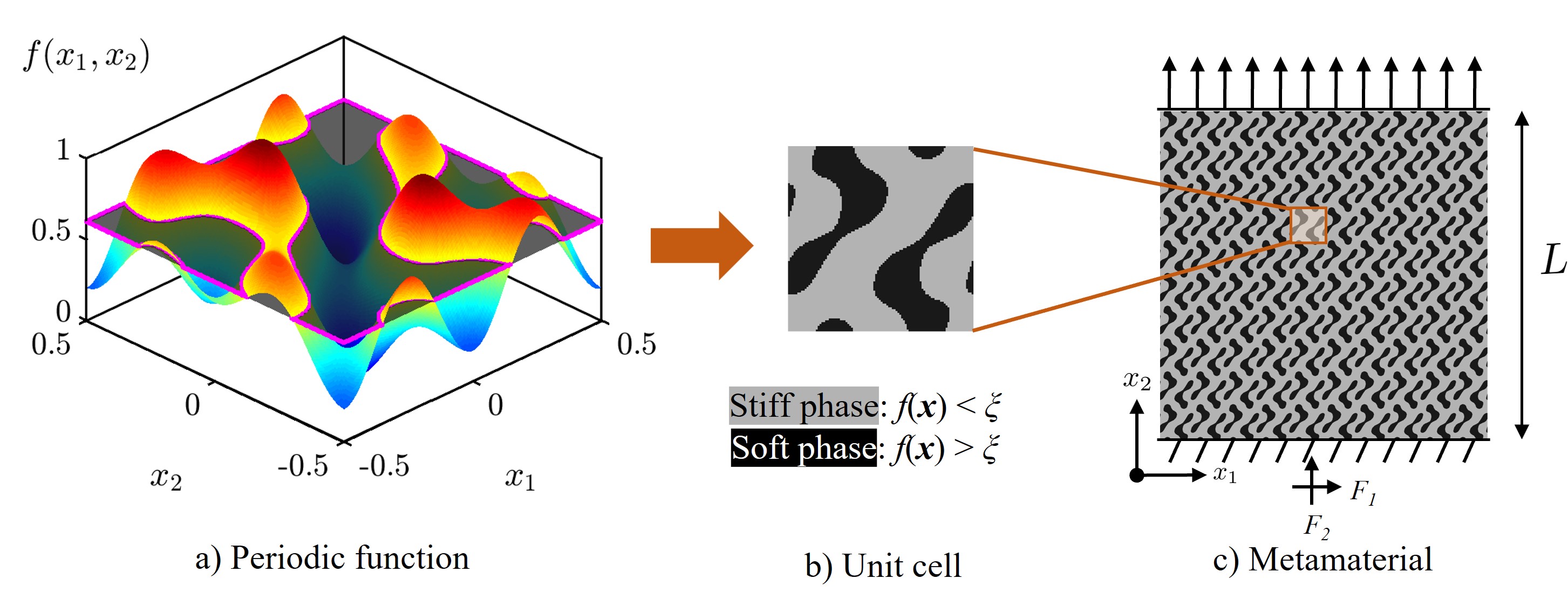}
		\caption{a) Design of an anisotropic unit cell geometry by thresholding a periodic function $f(x_1,x_2)$. b) A two-phase unit cell geometry consisting of a stiffer (gray) and a softer phase (black). c) A two-dimensional anisotropic metamaterial created by tessellating the unit cell geometry (shown in the inset) ten times along both $x_1$- and $x_2$- axes.}	\label{fig: 1 example anisotropic unit cell}
	\end{center}
\end{figure}

Without loss of generality, we study two-dimensional structured solids, obtained from finite periodic tessellation of square unit cells (\cref{fig: 1 example anisotropic unit cell})\footnote{Our methods are easily extendable to non-square unit cells.}. 
We focus on identifying the effective anisotropic material parameters of these composite assemblies, as linear elastic continua.


To design unit cells, we follow an approach inspired by Cahn's method of generating Gaussian random fields by superposing plane waves of fixed wavelength but random in phase and direction \citep{cahn_phase_1965,Soyarslan2018,Kumar2020}. 
We first define a function $f(x_1,x_2)$, as a linear superposition of cosine periodic functions: 
\begin{equation}
f(x_1,x_2) =  \sum_{m,n} A_{mn}\cos\left(2 \pi(m x_1+n x_2)\right)\label{eq: periodic functions}, \quad \forall (x_1,x_2) \in [-0.5,0.5], \quad \forall m,n \in [-3,-2,-1,0,1,2,3],
\end{equation}
where $m,n$ are spatial frequencies, and $A_{mn}$ are the corresponding cosine function weights.
The function is then thresholded at a value $\xi$, to generate a binary image which represents a unit cell, as shown in \cref{fig: 1 example anisotropic unit cell}, \textit{panels a, b}.
Each unit cell is pixelated and discretized with a 100 $\times$ 100 square mesh.
In this pixelated representation, the gray phase represents a stiffer material and the black phase represents a softer material (see \cref{subsec: exp}).

The periodicity is ensured from the choice of the cosine functions directly. 
We randomly sample the weights $A_{mn}$ and the threshold value $\xi$ to generate a small database of unit cells (about 100), from which we pick four unit cells to study in this paper. The four unit cells are chosen such that they are diverse in anisotropic properties and suitable for additive manufacturing  (see \cref{subsec: designofgeoms}).
{We consider a unit cell as suitable for manufacturing if the stiff phase is connected in the finite periodic tessellation with a minimum feature size of 5 pixels, matching the resolution of our chosen additive manufacturing approach.} 

A schematic of our setup is shown in \cref{fig: 1 example anisotropic unit cell}\textit{c}. 
A two-dimensional square anisotropic structured solid with $10 \times 10$ {unit cell} tessellation, {with side length \textit{L}}, is subjected to a displacement-controlled tension test. 
The boundary conditions are such that the bottom end is fixed, while a displacement of $\bm{u} = [0, u_p]^T$ is prescribed at the top end.
The reaction force components measured at the fixed end are denoted as $F_1, F_2$. 
\section{Virtual fields method for anisotropic metamaterials}
Many parameter identification methods rely on conducting multiple experiments, which are time consuming, complex and require specialized equipment.
To circumvent these drawbacks, we explore a material characterization method based on the VFM that solely relies on full-field displacements and net reaction force measurements from a single experimental test. In this section, after discussing the assumptions underlying the adoption of the VFM for metamaterials, we outline all the components of the proposed method. 
\label{sec: VFM}
\subsection{Basic assumptions}
\label{basic_assump}
The VFM \citep{grediac_principle_1989,grediac_virtual_2008,pierron_virtual_2012} exploits the weak formulation of linear momentum balance, i.e., the principle of virtual work, as a constraint on the material parameter space.
Since the full displacement field over the specimen and the net reaction forces at the specimen boundaries are known, testing the weak formulation for a suitable set of test functions (also known as virtual fields) results in a system of equations that can be solved for the unknown material parameters.
By choosing the test functions as not constant in space, the linear momentum balance is tested in different regions of the considered specimen domain.
As such, the VFM takes advantage of the local strain data, as opposed to global methods for parameter identification.

In the following, the VFM is used to characterize the mechanical behavior of metamaterials. 
However, it should be noted that -- due to the non-homogeneous nature of the metamaterials -- the application of identification methods based on full-field measurements is not trivial.
Full-field measurement techniques such as DIC measure the kinematic fields locally, i.e., at several points on the considered specimen surface.
The studied metamaterials are not expected to behave at these local points as their homogenized counterparts, especially when the number of repeating unit cells is low in comparison to the size of the specimen.
To give an example, in \cref{sec: synthetic data} the deformation of a heterogeneous metamaterial specimen will be compared to that of an equally-dimensioned homogeneous body, whose stiffness is set to the homogenized stiffness of the metamaterial.
Under the same loading conditions, the two specimens exhibit different local displacements, which is likely caused by local size effects and the different boundary conditions that are assumed during the loading of the macroscopic structure and the homogenization of the microscopic unit cell.
It is observed that deviations between the kinematic fields are predominant at the boundary and in particular at the corners of the domain.
This agrees with theoretical studies on heterogeneous metamaterials, which suggest the usage of non-local – e.g., higher-order strain-gradient based – theories as proposed by \citep{mindlin_first_1968}, to model size effects and wedge forces appearing at corners of non-homogeneous bodies \citep{fischer_isogeometric_2011,andreaus_numerical_2016,yang_verification_2021}.
Within this work, such theories are avoided for the sake of simplicity and to keep a reasonably low number of material parameters. Hence, the assumption is made that the global material behavior of the metamaterials can be characterized based on local kinematic measurements within a local constitutive theory.
As we will see later, this assumption will introduce errors in the identification procedure, which are, however, below a practically relevant level.
During the development of the VFM, we found that the locally measured kinematic data must be treated with care, especially at the boundary and the corners of the specimen.
We will later introduce specifically designed virtual fields that reduce the influence of data acquired at the specimen boundary and corners (see \cref{sec:choice_test_functions} for details).\footnote{We note at this point that reducing the influence of data acquired at the specimen boundary and corners may be beneficial not only when studying heterogeneous materials. Even for homogeneous specimens, the acquisition of kinematic data at the specimen boundary via DIC is known to be difficult.}

\subsection{Required data}
To identify the unknown parameters, the VFM needs diverse local strain data, i.e., strain fields that are not homogeneous.
Therefore, data that serve as input for the VFM are usually generated by testing complex specimen geometries under complex loading conditions.
For our purposes we will show that, due to the anisotropy of the material, a clamped square plate under uniaxial tension produces a sufficiently heterogeneous strain field.
We hence consider a displacement-controlled uniaxial tension experiment of a square-shaped specimen that consists of $n_c \times n_c$ repeating square unit cells of the considered metamaterial (\cref{fig: 1 example anisotropic unit cell}). 
At the fixed boundary of the specimen, a load cell measures the net reaction force. Further, the full-field deformation of the specimen is tracked through DIC, which measures the local displacements of the solid material. 
After preprocessing the data, the VFM takes as input the displacement measurements at the $(n_c+1) \times (n_c+1)$ unit cell corners and the net reaction forces.
A quadrilateral finite element mesh is generated such that each of the $n_c \times n_c$ elements corresponds to one unit cell and the element nodes correspond to the unit cell corners with experimentally known displacement values. The continuous displacement field $\bm{u}(\bm{x})$ is hence approximated by
\begin{align}
\label{eq:displacement_function}
\bm{u}(\bm{x}) = \sum_{a=1}^{n_n} N^a(\bm{x})\bm{u}^a,
\end{align}
where $n_n = (n_c+1)^2$ denotes the number of nodes in the finite element mesh and $\bm{u}^a$ are the known nodal displacements, while $N^a(\bm{x})$ are the standard ansatz functions of bilinear quadrilateral finite elements. 
The infinitesimal strain field is then obtained as the symmetric gradient of the displacement field, i.e., $\bm{\varepsilon}(\bm{x}) = \frac{1}{2} \left(\nabla\bm{u}(\bm{x}) + \left(\nabla\bm{u}(\bm{x})\right)^T\right)$.

\subsection{Weak formulation of linear momentum balance}
We denote the specimen domain and its boundary as $\Omega$ and $\partial \Omega$, respectively, and the surface traction force acting on $\partial \Omega$ as $\bm{t}$.
Assuming no inertia and body forces, the weak form of linear momentum balance reads
\begin{align}
\int_\Omega \bm{\sigma}(\bm{x})\colon\nabla\bm{v}(\bm{x}) \ \mathrm{d} A - \int_{\partial\Omega} \bm{t} \cdot \bm{v}(\bm{x}) \ \mathrm{d} s = 0,
\end{align}
which has to hold true for all admissible, i.e., sufficiently regular, test functions $\bm{v}(\bm{x})$. Note that we are not introducing the classical distinction between Dirichlet and Neumann portions of the boundary; accordingly, we are not requiring admissible test functions to vanish anywhere.


\subsection{Discretization}
The weak form of linear momentum balance has to hold true for any chosen set of admissible test functions.
Here, we adopt the standard (Bubnov-Galerkin) approach and express the test functions as a linear combination of the same shape functions $N^a(\bm{x})$ used to interpolate the displacement data
\begin{align}
\label{eq:test_function}
\bm{v}(\bm{x}) = \sum_{a=1}^{n_n} N^a(\bm{x})\bm{v}^a.
\end{align}
Inserting the test function ansatz into the weak form of linear momentum balance results in
\begin{align}
\label{eq:force_balance}
\sum_{a=1}^{n_n}\bm{v}^a\cdot\left[ \underbrace{\int_\Omega \bm{\sigma}\nabla N^a(\bm{x}) \ \mathrm{d} A}_{\bm{F}_{\text{int}}^{a}} - \underbrace{\int_{\partial\Omega} \bm{t} N^a(\bm{x}) \ \mathrm{d} S}_{\bm{F}_{\text{ext}}^{a}}\right] = 0,
\end{align}
where the first and second integral are the nodal internal forces $\bm{F}_{\text{int}}^{a}$ and nodal external forces $\bm{F}_{\text{ext}}^{a}$, respectively.
By employing the constitutive relation \JB{\cref{eq: ConstIndicialSmall}}, the nodal internal forces may be written as
\begin{align}
\label{eq:F_int}
\bm{F}_{\text{int}}^{a}
&= \int_\Omega \bm{\sigma}\nabla N^a \ \mathrm{d} A, \notag\\
&= \int_\Omega  
\begin{bmatrix}
\sigma_{{1}}N_{,x}^a + \sigma_{{6}}N_{,y}^a\\
\sigma_{{6}}N_{,x}^a + \sigma_{{2}}N_{,y}^a\\
\end{bmatrix}
\ \mathrm{d} A, \notag\\
&= \int_\Omega  
\begin{bmatrix}
C_{11}\varepsilon_{{1}}N_{,x}^a + C_{12}\varepsilon_{{2}}N_{,x}^a + 2C_{16}\varepsilon_{{6}}N_{,x}^a
+ C_{16}\varepsilon_{{1}}N_{,y}^a + C_{26}\varepsilon_{{2}}N_{,y}^a + 2C_{66}\varepsilon_{{6}}N_{,y}^a\\
C_{16}\varepsilon_{{1}}N_{,x}^a + C_{26}\varepsilon_{{2}}N_{,x}^a + 2C_{66}\varepsilon_{{6}}N_{,x}^a
+ C_{12}\varepsilon_{{1}}N_{,y}^a + C_{22}\varepsilon_{{2}}N_{,y}^a + 2C_{26}\varepsilon_{{6}}N_{,y}^a\\
\end{bmatrix}
\ \mathrm{d} A, \notag\\
&= \int_\Omega  
\begin{bmatrix}
\varepsilon_{{1}}N_{,x}^a &
\varepsilon_{{2}}N_{,x}^a &
0&
2\varepsilon_{{6}}N_{,x}^a + \varepsilon_{{1}}N_{,y}^a&
\varepsilon_{{2}}N_{,y}^a&
2\varepsilon_{{6}}N_{,y}^a\\
0 &
\varepsilon_{{1}}N_{,y}^a &
\varepsilon_{{2}}N_{,y}^a &
\varepsilon_{{1}}N_{,x}^a &
\varepsilon_{{2}}N_{,x}^a + 2\varepsilon_{{6}}N_{,y}^a &
2\varepsilon_{{6}}N_{,x}^a\\
\end{bmatrix}
\ \mathrm{d} A \
\bm{C}_{\text{vec}},
\end{align}
where the elasticity tensor parameters $\bm{C}_{\text{vec}} = [ C_{11} \ C_{12} \ C_{22} \ C_{16} \ C_{26} \ C_{66}]^T$ are assumed to be constant in space. 

\subsection{Choice of test functions}
\label{sec:choice_test_functions}
Choosing a test function in the form of \eqref{eq:test_function} and evaluating \eqref{eq:force_balance} results in two linear equations with the material parameters as unknowns. As the weak linear momentum balance has to hold true for any test function this provides an infinite supply of linear equations. Hence, the problem at hand is overdetermined and different choices of test functions will yield different solutions for the unknown material parameters. 

As discussed in \cref{basic_assump},
the deformation of a heterogeneous specimen and that of its homogenized counterpart under the same loading conditions are locally different, a phenomenon that is best observed at the boundary and at the corners of the specimen where local effects are especially pronounced.
In the following, this special characteristic of the problem at hand motivates a special choice of the test functions that avoids evaluations of the linear momentum balance in the boundary regions of the specimen.

First, we define test functions that are constant at the nodes corresponding to one finite element, i.e., one unit cell, and zero at all other nodes. 
To this end, we define $\mathcal{C}=\{1,\dots,n_c^2\}$ as the set of all unit cells and $\mathcal{D}^c$ as the set of all nodes corresponding to the unit cell $c\in\mathcal{C}$, and define a set of test functions as
\begin{align}
\label{eq:test_functions_unit_cell}
\mathcal{V} = \left\{\bm{v}(\bm{x}) = \frac{1}{n_{nc}} \sum_{a\in\mathcal{D}^c} N^a(\bm{x})\bm{e}_i \ | \ c \in \mathcal{C}, \ i\in\{1,2\} \right\},
\end{align}
where $\bm{e}_i$ are the unit vectors in the corresponding $x$- and $y$-direction. Note that the test functions are normalized by dividing by the number of nodes corresponding to the unit cell $n_{nc}$ (equal to 4 in our case).

\begin{figure}[!htb]
	\begin{center}
		\centering 
		\includegraphics[width=0.8\textwidth]{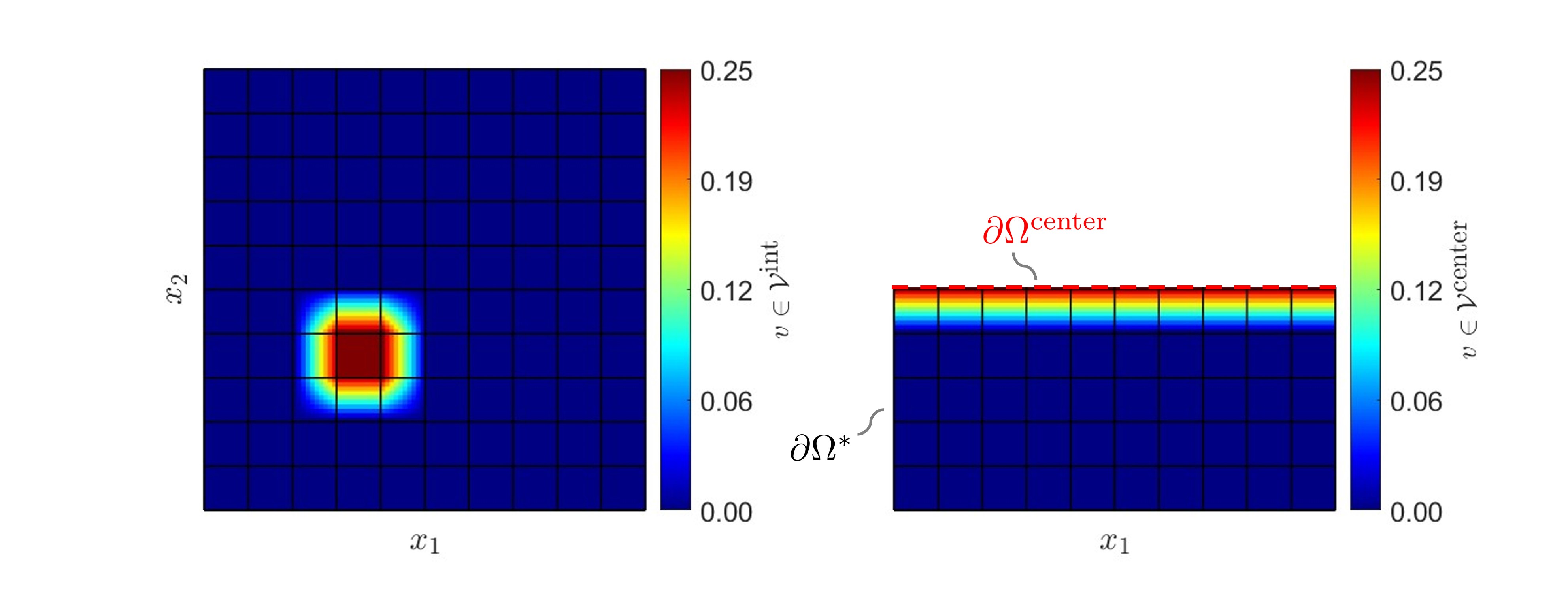}
		\caption{Non-zero component of a virtual field in $\mathcal{V}^{\text{int}}$ (left) and non-zero component of a virtual field in $\mathcal{V}^{\text{center}}$ (right).}
		\label{fig:virtual_fields}
	\end{center}
\end{figure}

Using the test functions in $\mathcal{V}$ to test weak linear momentum balance would cause two problems.
First, at elements adjacent to the loaded and to the restrained portions of the boundary, the external force contributions $\bm{F}_{\text{ext}}^{a}$ in \eqref{eq:force_balance} are unknown, leading to equations that could not be solved for the unknown material parameters.
And second, we want to avoid using data at the specimen boundary due to the reasons discussed earlier. Therefore, we modify \eqref{eq:test_functions_unit_cell} such that
\begin{align}
\label{eq:test_functions_int}
\mathcal{V}^{\text{int}} = \left\{\bm{v}(\bm{x}) = \frac{1}{n_{nc}} \sum_{a\in\mathcal{D}^c} N^a(\bm{x})\bm{e}_i \ | \ c \in \mathcal{C}^{\text{int}}, \ i\in\{1,2\} \right\},
\end{align}
where $\mathcal{C}^{\text{int}}\subset\mathcal{C}$ denotes a reduced set of unit cells that does not include unit cells close to the boundary. We found that ignoring two rows of unit cells at the top and bottom boundary as well as two columns of unit cells at the left and right boundary are a good compromise, and we kept this choice constant throughout all tests. As the fields in $\mathcal{V}^{\text{int}}$ depend on $\bm{e}_i$, each field is zero in either $x$- or $y$-direction. The non-zero component of an exemplary virtual field in $\mathcal{V}^{\text{int}}$ is shown in \cref{fig:virtual_fields} (left).

Evaluating \cref{eq:force_balance} for this set of functions leads to
\begin{align}
\label{eq:eqns_int}
\frac{1}{n_{nc}} \sum_{a\in\mathcal{D}^c} \bm{F}_{\text{int}}^{a} = \bm{0}, \quad \forall c \in \mathcal{C}^{\text{int}}.
\end{align}
Hence, this choice of virtual fields can be interpreted physically as enforcing that the sum of internal forces over one unit cell should vanish.


Equations \eqref{eq:eqns_int} are not sufficient to identify the unknown material parameters, as the trivial solution $\bm{C}_{\text{vec}} = \bm{0}$ fulfills \eqref{eq:eqns_int}. 
To obtain a well-posed problem, the measured reaction forces need to be incorporated. 
At the same time, we want to avoid using displacement data at the specimen boundary. 
Therefore, we consider the free-body diagram of the lower half of the domain as depicted in \cref{fig:virtual_fields} (right).
Denoting the half-body domain as $\Omega^*=\{\bm{x} \ | \ 0 \leq x_1 \leq L, \ \ 0 \leq x_2 \leq \frac{L}{2} \}$ and its boundary as $\partial\Omega^*$, the weak form of linear momentum balance for this domain reads
\begin{align}
\int_{\Omega^*} \bm{\sigma}(\bm{x})\colon\nabla\bm{v}(\bm{x}) \ \mathrm{d} A - \int_{\partial\Omega^*} \bm{t} \cdot \bm{v}(\bm{x}) \ \mathrm{d} s = 0.
\end{align}
Inserting the test function ansatz leads to
\begin{align}
\label{eq:force_balance*}
\sum_{a=1}^{n_n}\bm{v}^a\cdot\left[ \underbrace{\int_{\Omega^*} \bm{\sigma}\nabla N^a(\bm{x}) \ \mathrm{d} A}_{\bm{F}_{\text{int}}^{*a}} - \underbrace{\int_{\partial\Omega^*} \bm{t} N^a(\bm{x}) \ \mathrm{d} S}_{\bm{F}_{\text{ext}}^{*a}}\right] = 0.
\end{align}
We define $\mathcal{D}^{\text{center}} =  \{a \ | \ y^a = \frac{L}{2}\}$ as the set of nodes in the center of the specimen. If the tessellated geometry consists of an odd number of unit cells in each spatial direction, i.e., there are no nodes at $y^a = \frac{L}{2}$, we consider instead $\mathcal{D}^{\text{center}} =  \{a \ | \ y^a = \frac{L}{2} + \frac{L}{2 n_c}\}$. We choose a set of virtual fields $\mathcal{V}^{\text{center}}$ that are constant along $\mathcal{D}^{\text{center}}$ and zero at all other nodes
\begin{align}
\mathcal{V}^{\text{center}} = \left\{\bm{v}(\bm{x}) =  \frac{1}{n_{nc}} \sum_{a \in \mathcal{D}^{\text{center}}} N^a(\bm{x})\bm{e}_i \ | \ i\in\{1,2\} \right\}.
\end{align}
Evaluating \eqref{eq:force_balance*} for these particularly chosen test functions results in
\begin{align}
\label{eq:eqns_center}
\sum_{a \in \mathcal{D}^{\text{center}}} \bm{F}_{\text{int}}^{*a} = \int_{\partial\Omega^{\text{center}}} \bm{t} \ \mathrm{d} S = \bm{R},
\end{align}
where $\partial\Omega^{\text{center}}$ is the top boundary of $\Omega^*$. Note that due to the specific choice of the test functions, the surface integral simplifies in such a way that it equals the global reaction force $\bm{R}$, meaning that the sum of the internal forces at $\partial\Omega^{\text{center}}$ must equal the net reaction force.

\subsection{Deterministic parameter identification}
\label{sec: VFM param}
After choosing the virtual fields and considering \eqref{eq:F_int}, the linear equations in \eqref{eq:eqns_int} can be assembled in a system of equations
\begin{equation}
\label{eq:A_int}
\bm{A}^{\text{int}}\bm{C}_{\text{vec}}=\bm{0},
\end{equation}
and the linear equations in \eqref{eq:eqns_center} can be rewritten as
\begin{equation}
\label{eq:A_center}
\bm{A}^{\text{center}}\bm{C}_{\text{vec}}=\bm{R},
\end{equation}
where $\bm{A}^{\text{int}}$ and $\bm{A}^{\text{center}}$ are in general non-symmetric matrices. The system formed by the linear equations \eqref{eq:A_int} and \eqref{eq:A_center} is overdetermined, i.e., it consists of more equations than unknown parameters. 
Assuming that the equations in the overdetermined system are not linearly dependent (which is a valid assumption as every equation is perturbed by noise when considering experimental data), there is no unique solution that satisfies all equations.
Instead, we obtain an approximate solution of the overdetermined system by minimizing the sum of squared residuals
\begin{equation}
\label{eq:min_problem1}
\bm{C}_{\text{vec}}^{\text{opt}} = \argminWithArgs_{\bm{C}_{\text{vec}}} \left( \|\bm{A}^{\text{int}}\bm{C}_{\text{vec}}\|^2 + \lambda_r \|\bm{A}^{\text{center}}\bm{C}_{\text{vec}}-\bm{R}\|^2\right),
\end{equation}
where $\|\cdot\|$ is the Euclidean norm and $\lambda_r > 0$ is a weighting parameter that scales the different contributions to the minimization problem.
As there are less equations in the system \eqref{eq:A_center} than in \eqref{eq:A_int}, the weighting parameter should be chosen sufficiently larger than one ($\lambda_r>>1$).
Following previous works \cite{flaschel_unsupervised_2021, flaschel_discovering_2022, flaschel_automated_2023}, we choose $\lambda_r = 100$ and keep it constant throughout this work. 
Based on our experience, the choice of $\lambda_r$ is not crucial for the success of the method (see also \cite{joshi_bayesian-euclid_2022,thakolkaran_nn-euclid_2022,marino_automated_2023}).
The necessary condition for a minimum is
\begin{equation}
\label{eq:min_problem1_nc}
\bar{\bm{A}}\bm{C}_{\text{vec}}^{\text{opt}}=\bar{\bm{R}},
\qquad \text{with}
\qquad \bar{\bm{A}} = \left(\bm{A}^{\text{int}}\right)^T\bm{A}^{\text{int}} + \lambda_r \left(\bm{A}^{\text{center}}\right)^T\bm{A}^{\text{center}}
,~ \bar{\bm{R}} = \lambda_r \left(\bm{A}^{\text{center}}\right)^T\bm{R},
\end{equation}
which leads to a determined system of equations that can be solved for $\bm{C}_{\text{vec}}^{\text{opt}}$.
The minimization problem in \cref{eq:min_problem1} can alternatively be written as
\begin{equation}
\label{eq:min_problem2}
\bm{C}_{\text{vec}}^{\text{opt}} = \argminWithArgs_{\bm{C}_{\text{vec}}} \|\bm{A}\bm{C}_{\text{vec}} - \bm{B}\|^2,
\end{equation}
where we have defined
\begin{equation}
    \bm{A} = \begin{bmatrix}
    \bm{A}^{\text{int}} \\
    \sqrt{\lambda_r}\bm{A}^{\text{center}} \\
    \end{bmatrix}, \qquad
    \bm{B} = \begin{bmatrix}
    \bm{0} \\
    \sqrt{\lambda_r}\bm{R} \\
    \end{bmatrix}.
\end{equation}
The necessary condition for a minimum then reads
\begin{equation}
\label{eq:min_problem2_nc}
\bm{A}^T\bm{A}\bm{C}_{\text{vec}}^{\text{opt}}=\bm{A}^T\bm{B}.
\end{equation}

\subsection{Bayesian inference}
\label{sec:Bayesian_inference}
Besides the previously introduced deterministic approach, we further study the problem from a stochastic perspective.
To this end, we construct a Bayesian linear regression model, for which we assume no intercept and a diffuse prior, as implemented in the Matlab\,\textsuperscript{\tiny\textregistered} built-in function \textit{bayeslm}.

We denote the number of rows in $\bm{A}$ as $n_{eq}$ and we define $\bm{A}_i$ with $i\in\{1,\dots,n_{eq}\}$ as the $i^\text{th}$ row of $\bm{A}$.
For each equation in the overdetermined system of equations $\bm{A}\bm{C}_{\text{vec}} = \bm{B}$, we assume the likelihood of obtaining $B_i$ as a Gaussian likelihood with mean $\bm{A}_i \cdot \bm{C}_{\text{vec}}$ and standard deviation $\sigma>0$, i.e.,
\begin{equation}
p(B_i|\bm{A}_i,\bm{C}_{\text{vec}},\sigma^2) 
= \frac{1}{\sqrt{2\pi\sigma^2}} \exp \left[ - \frac{\left( B_i - \bm{A}_i \cdot \bm{C}_{\text{vec}} \right)^2}{2\sigma^2} \right],
\end{equation}
where $\bm{C}_{\text{vec}}$ and $\sigma^2$ are treated as random variables.
Assuming further that the likelihoods are conditionally independent, we define the joint likelihood as
\begin{equation}
    p(\bm{B}|\bm{A},\bm{C}_{\text{vec}},\sigma^2) = \prod_{i=1}^{n_{eq}} p_i(B_i|\bm{A}_i,\bm{C}_{\text{vec}},\sigma^2).
\end{equation}
Assuming here a diffuse prior for the joint prior distribution of $\bm{C}_{\text{vec}}$ and $\sigma^2$, i.e.,
\begin{equation}
    p(\bm{C}_{\text{vec}},\sigma^2) \propto \frac{1}{\sigma^2},
\end{equation}
the marginal posterior distributions of $\bm{C}_{\text{vec}}$ and $\sigma^2$ are analytically tractable and implemented in the Matlab\,\textsuperscript{\tiny\textregistered} function \textit{bayeslm}.

\section{Data acquisition}
\label{sec: data aq}
In this section, we first discuss the unit cell geometries considered for identification of the material parameters. Then, we describe our numerical and experimental data acquisition methods, including details on fabrication, experimental setup, testing and DIC.
\subsection{Design and choice of unit cell geometries}
\label{subsec: designofgeoms}
We pick four unit cells with distinct/diverse effective stiffness tensor parameters (all with six non-zero stiffness parameters). 
\cref{tab:geometries} shows the unit cells along with their symmetry class and homogenized stiffness tensor.
Geometry \#1 has $C_{22}$ as the largest stiffness parameter with {$C_{16}$ almost comparable to $C_{12}$} 
and $C_{26} > C_{16}$.  
While geometry \#2 has $C_{11}$ as the largest stiffness parameter with $C_{16} > C_{26}$, geometry \#3 has negative values for all of the off-diagonal parameters. 
Geometry \#4 has four independent stiffness parameters with $C_{66}$ as one of the largest values among other stiffness parameters, along with $C_{11} = C_{22}$ and $C_{16} = C_{26}$.
The fill fraction of the stiff phase for all the unit cells lies between 60 and 70 \%.

\newcolumntype{P}[1]{>{\centering\arraybackslash}p{#1}} 
\newcolumntype{M}[1]{>{\centering\arraybackslash}m{#1}} 
\newcolumntype{R}[1]{>{\raggedleft\arraybackslash}p{#1}} 
\begin{table}[hbtp]
\centering
\begin{tabular} {M{2.5cm} M{2.5cm} M{5cm} M{2.5cm}}
\toprule
\textbf{Unit Cell Geometry} & \textbf{Name} & \textbf{Homogenized Stiffness Tensor} $(\bm{C}^\text{H})$ [MPa] & \textbf{Elastic Symmetry Class}\\%
\midrule
\includegraphics[width =0.1\textwidth]{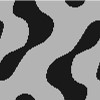}&  Geometry \#1  & ${\left[\begin{array}{rrr} 131.62&61.98&63.58\\61.98&198.38&83.87\\63.58&83.87&95.30\end{array}\right]}$ & $Z_2$\\%
\includegraphics[width =0.1\textwidth]{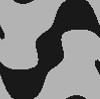} & Geometry \#2 & ${\left[\begin{array}{rrr}  127.14 & 59.50 & 73.42 \\  59.50 & 105.83 & 55.16 \\ 73.42 & 55.16 & 110.15\end{array}\right]}$ & $Z_2$ \\%
\includegraphics[width =0.1\textwidth]{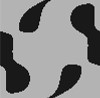} & Geometry \#3 & ${\left[\begin{array}{rrr} 44.70 & -9.42 & -12.52 \\  -9.42 & 107.19 & -20.71\\ -12.52 & -20.71 & 105.35\end{array}\right]}$ & $Z_2$\\%
\includegraphics[width =0.1\textwidth]{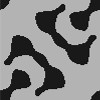} & Geometry \#4 & ${\left[\begin{array}{rrr} 65.74 & 40.36 & 18.95 \\ 40.36&65.74&18.95\\18.95&18.95&86.47\end{array}\right]}$ & $D_{2}$\\%
\bottomrule
\end{tabular}
\caption{Unit cell geometries considered in this study along with their mechanical and symmetry properties.} 
\label{tab:geometries}
\end{table}

\subsection{Numerical data generation}
\textit{Simulations:} We use synthetic data generated using the FEM to verify our methods and aid our analysis before performing the experiments. 
Each pixel is discretized using a four-node plane-stress bilinear quadrilateral element.
For tessellation, we vary the number of unit cells $n_c$ between 5 and 25. 

\textit{Homogenization:} We compute the effective mechanical properties of the unit cells using the theory of homogenization implemented using the FEM (as in \cite{andreassen_how_2014}). 

\subsection{Experimental data generation}
\subsubsection{Fabrication} 
\label{subsec: exp}
As specimens with a large number of unit cells are difficult to fabricate, we here pick 10 $\times$ 10 tessellations to perform experimental validations.
We use a commercial multi-material polyjet technology based 3D printer, Stratasys Objet500 Connex, to fabricate all the specimens.
The dimensions of the specimen are 75 × 75 × 5 mm excluding the portion that goes into the grips.
We use Stratasys' proprietary material DM8530 for the stiff phase and TangoBlack for the soft phase.
The material properties (DM8530: Young’s modulus \textit{E} = 1000 $\pm$ 90 MPa and Poisson’s ratio $\nu$= 0.35, TangoBlack: Young’s modulus \textit{E} =  0.7 MPa and Poisson’s ratio $\nu$= 0.49) are experimentally measured following the ASTM D638-14 standard test method and the same values are used in the numerical computations. 

\subsubsection{Experimental setup and testing} 
\label{subsection: experiments}
\begin{figure}[!htb]
	\begin{center}
		\centering%
        \includegraphics[width =0.95\textwidth]{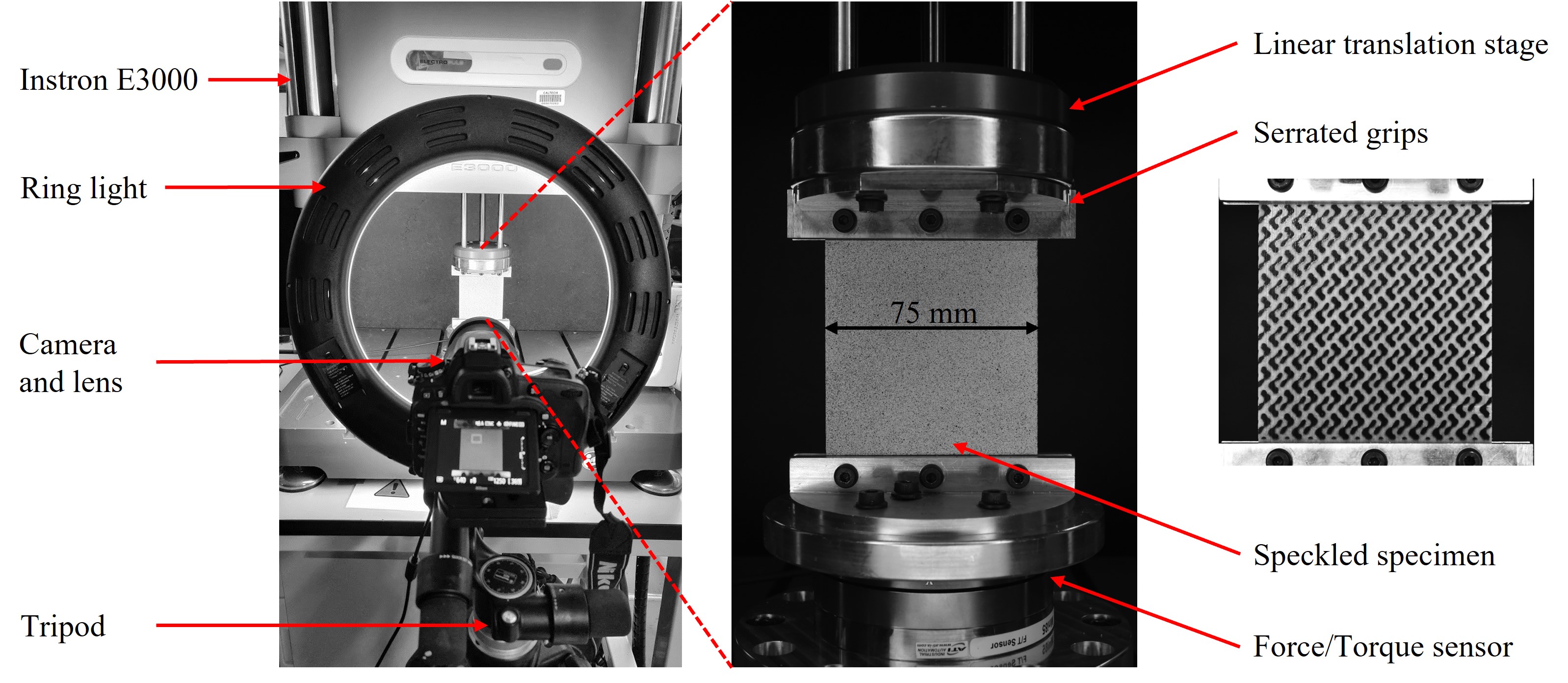}
		\caption{Experimental setup for displacement-controlled uniaxial testing of an anisotropic metamaterial.}
		\label{fig: experimentalsetup}
	\end{center}
\end{figure}

We subject the additively manufactured specimens to displacement-controlled tension tests using a universal testing machine, Instron E3000, mounted with a multi-axis force-torque sensor (ATI Mini85) as shown in \cref{fig: experimentalsetup}. 
The force-torque sensor is acquired from ATI Industrial Automation.
We apply a vertical displacement of 1.5 mm at the top boundary at a rate of 0.5 mm/min resulting in a global axial strain of $\tilde{\varepsilon}_{22}=0.02$ and a global strain rate of $ 1.1 \times 10^{-4}$s$^{-1}$. 
Custom designed grips are fabricated out of aluminum and are serrated to hold the specimens firmly and prevent any lateral slipping.
We use the same strain rate while measuring the constitutive material properties of the individual phases.

We use DIC, an image-based optical technique, to measure the full-field displacements \citep{sutton_image_2009}. 
We capture images at a frequency of 1 Hz using a Nikon D750 camera equipped with a Nikon AF-S NIKKOR 24-120mm f/4G ED VR zoom lens.
We use manual mode at an exposure rate of 1/640 sec, an ISO setting of 1250 and an aperture setting of F8. 
The camera has a 6016 by 4016 square pixel resolution and the region of interest we studied is about 3060 by 3060 pixels.
We place a ring light between the 
additively manufactured specimen and the camera to illuminate the surface uniformly and we place the camera lens at a distance of about 35-40 cm from the specimen plane.

\subsubsection{Digital image correlation}
Given a reference image $f$ and and a deformed image $g$, {the} correlation algorithm aims at minimizing the sum of squared differences over the considered domain $\Omega$
\begin{equation}
    \mathcal{T} = \int_{\Omega} \left(g(\bm{x}+\bm{u}(\bm{x}))-f(\bm{x})\right)^2 d\bm{x}, \label{eq: DIC equation}
\end{equation}
where $\bm{x}$ is the position in the reference image and $\bm{u}(\bm{x})$ is the displacement field which is interpolated as
\begin{equation}
    \bm{u}(\bm{x}) = \sum u_n \bm{\phi}_n(\bm{x}),
\end{equation}
where $\bm{\phi}_n$ are a set of shape functions and $u_n$ the associated degrees of freedom. 
There are two approaches to determine the unknowns $u_n$, \textit{local DIC} and \textit{global DIC} \citep{hild_comparison_2012}. 
In the \textit{local} approach, the region of interest ($\Omega$) is divided into several sub-images known as subsets and the mean displacement of each subset is computed independently while minimizing the objective \cref{eq: DIC equation}. 
In the \textit{global} approach, shape functions defined through a finite element mesh over the whole region of interest are used \citep{besnard_finite-element_2006}. 
The \textit{global} approach assumes continuity of displacements over the entire region of interest which is well suited when the structure is heterogeneous.
Moreover, the \textit{global} approach provides the displacement information at the boundaries, which is hard to obtain using the \textit{local} approach.
The displacement data at the boundaries are an important input for the VFM.
{Hence, we follow the \textit{global} approach to perform the correlation in this study.}

We perform DIC using piece-wise linear shape functions defined on a triangular mesh to compute the displacements (as in \cite{agnelli_systematic_2021}). 
We choose an edge length of 18 pixels ($\sim$ 0.44 mm) to construct the triangular mesh.
We observe a noise floor of the order of 0.04 mm in the displacement data which is obtained from correlation performed on static images.
The data provided by the DIC correspond to the nodes that might not {always} align with the unit cell corners. 
To obtain the displacements of the unit cell corners, we further average the displacement data from the nodes that fall within 1 mm radius of a unit cell corner. 

\section{Results and discussion}
\label{sec: rd}
In this section, we discuss the data generation from both numerical simulations and experiments. Afterwards, we apply the proposed deterministic parameter identification method to the data and discuss the results. Finally, at the end of the section, we apply the Bayesian method to the data.

\subsection{Generation of full-field displacement data}
\subsubsection{Synthetic data}
\label{sec: synthetic data}

\begin{figure}[!htb]
	\begin{center}
		\centering 
  \includegraphics[width=\textwidth]{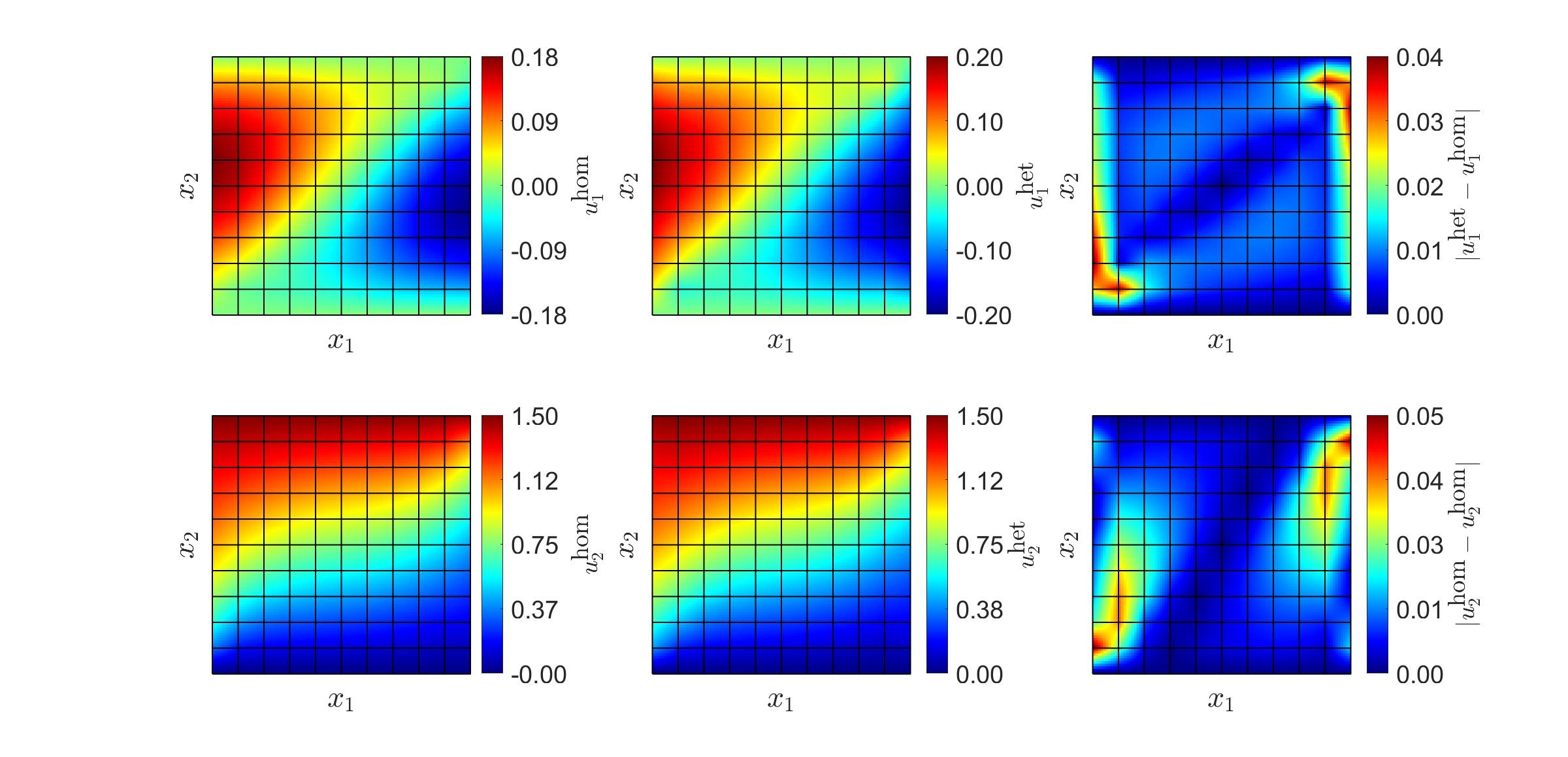}
		\caption{Comparison between the displacement fields obtained from finite element simulations of a homogeneous specimen (left) and a heterogeneous structure made of geometry \#1 (center). For the homogeneous specimen a finite element simulation using $10 \times 10$ bilinear quadrilateral elements was executed. The heterogeneous specimen was simulated using $1000 \times 1000$ bilinear quadrilateral elements. Afterwards, the displacement data at the unit cell corners were extracted and interpolated with a bilinear polynomial for each unit cell, to allow for a comparison with the homogeneous specimen. The difference between the fields is shown on the right.}
		\label{fig:displacement_HOMvsHET_geom01}
	\end{center}
\end{figure}

In the following, we investigate the synthetically generated displacement data for a heterogeneous structure in comparison to the computed displacement field of a homogeneous body, whose stiffness is equal to the homogenized stiffness of the heterogeneous structure. 
To simulate the displacement of a homogeneous body, we assume a $10 \times 10$ bilinear quadrilateral finite element mesh. The displacement of the heterogeneous body is computed on a much finer mesh with $1000 \times 1000$ elements.
To allow for a comparison with the displacement field of the homogeneous body, the computed displacements at the unit cell corners of the heterogeneous body (i.e. the data of interest for the VFM) are extracted and interpolated with a bilinear polynomial for each unit cell.
It can be seen in \cref{fig:displacement_HOMvsHET_geom01} that there is a good qualitative agreement between the two displacement fields {for geometry \#1} (see \cref{fig:displacement_HOMvsHET_geom02}, \cref{fig:displacement_HOMvsHET_geom03} and \cref{fig:displacement_HOMvsHET_geom04} for the other geometries).
However, there are quantitative differences due to local effects in the heterogeneous structure, which appear to be dominant at the boundary and corners of the specimen. 

\subsubsection{Comparison between experimental and synthetic data}

\begin{figure}[!htb]
	\begin{center}
		\centering 
		\includegraphics[width =0.99\textwidth]{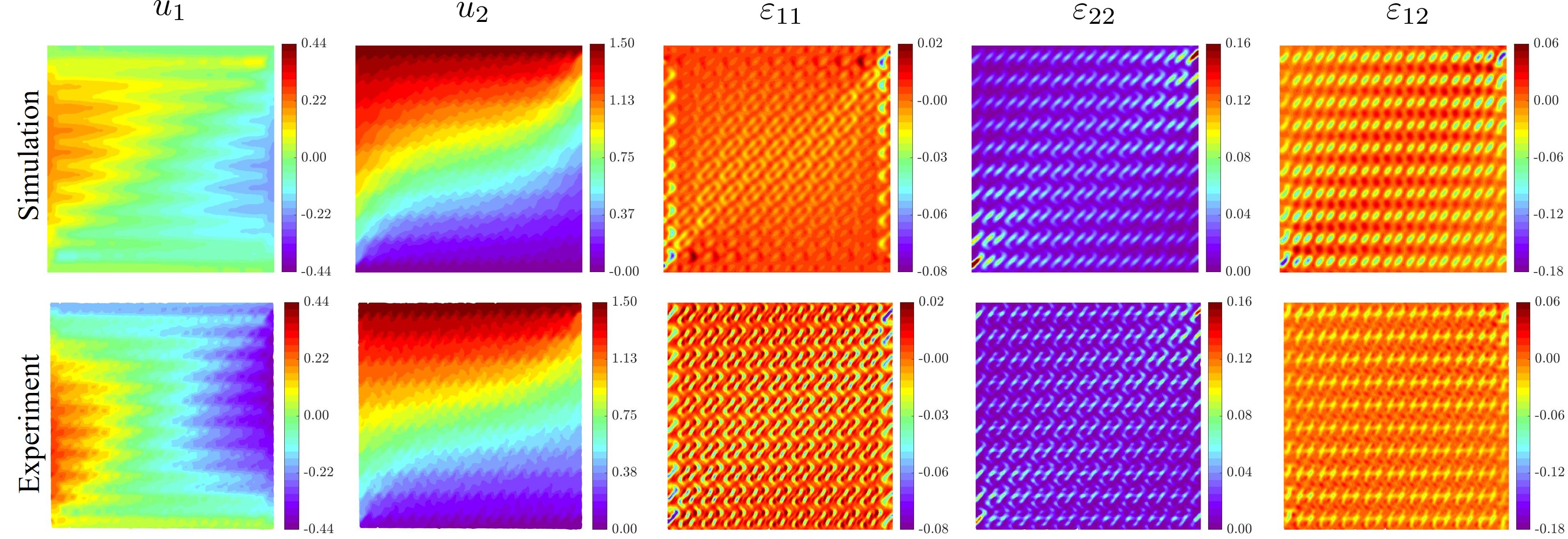}
		\caption{Comparison between numerical (top) and experimentally measured (bottom) full-field displacement and strain field data for the 10 unit cell tessellation of geometry \#1 subjected to displacement-controlled uniaxial tension test.} \label{fig: full field comparison geom01}
	\end{center}
\end{figure}

In \cref{fig: full field comparison geom01}, we compare the full-field displacement and strain fields between the numerical and experimental data on the heterogeneous structure for geometry \#1.
(See \cref{fig: full field comparison geom02,fig: full field comparison geom03,fig: full field comparison geom04} for the other geometries).
We observe very good agreement between the numerical and experimental data, especially for the variables $u_2, \varepsilon_{22}$. However, the experimentally measured $u_1$ appears to be slightly higher than the numerical data, by about 0.1 mm, for all the geometries. 
Also the two $\varepsilon_{11}$ fields are in good qualitative agreement, but experimental strains are larger. 
As expected, most of the strain is localized in the softer phase, although the applied global strain ($\tilde{\varepsilon}_{22}$) is 0.02. 

Further, a comparison of the displacement fields after postprocessing the synthetic and experimental data, i.e., after extracting and interpolating the displacements at the unit cell corners for all the geometries \sout{{for geometry \#2}} are shown in \cref{fig:displacement_FEMvsEXP_geom01}, \cref{fig:displacement_FEMvsEXP_geom02}, \cref{fig:displacement_FEMvsEXP_geom03}, and \cref{fig:displacement_FEMvsEXP_geom04}.
All displacements are in good agreement. 
An exception is observed for geometry \#3 (see \cref{fig: full field comparison geom03} and \cref{fig:displacement_FEMvsEXP_geom03}), for which the experimentally measured horizontal displacement $u_1$ does not compare well to the corresponding finite element results. 
The unit cell architecture of geometry \#3 leads to highly nonlinear mechanical behavior (see \cref{fig: loaddata}), which is not captured well in the simulations.
\subsection{Parameter identification based on synthetic data}
\label{sec: numerical convergence}
Since the homogenization theory assumes length scale separation and periodic boundary conditions in identifying the effective material parameters, it is important to understand the continuum behavior of the heterogeneous structures as the number of unit cells change. 
For this, we apply the VFM described in \cref{sec: VFM param} on the synthetic data to identify material parameters as the number of unit cells are varied.  
Further, we also use synthetic data to identify parameters using multiple tests (as in the conventional approach). 
A discussion on this conventional approach is provided in \ref{sec: multiple tests}.
The relative error is defined as
\begin{equation}
    \text{LSE}_{\|\cdot\|_{2}}=\frac{\left\|\bm{C}^{\mathrm{H}}_\text{vec}-\bm{C}^{\mathrm{M}}_\text{vec}\right\|_{2}}{\left\|\bm{C}^\mathrm{H}_\text{vec}\right\|_{2}}  \quad \text { with } \bm{C}_\text{vec} \in \mathbb{R}^{6}, 
    \label{eq: Cmat pred error}
\end{equation}
where $\bm{C}^{\mathrm{H}}_\text{vec}$ is the vectorized homogenized stiffness tensor obtained from computational homogenization and $\bm{C}^{\mathrm{M}}_\text{vec}$ is the vectorized stiffness tensor identified using the VFM and the conventional methods. 

\begin{figure}[!htb]
    \begin{center}
    \includegraphics[width=0.8\textwidth]{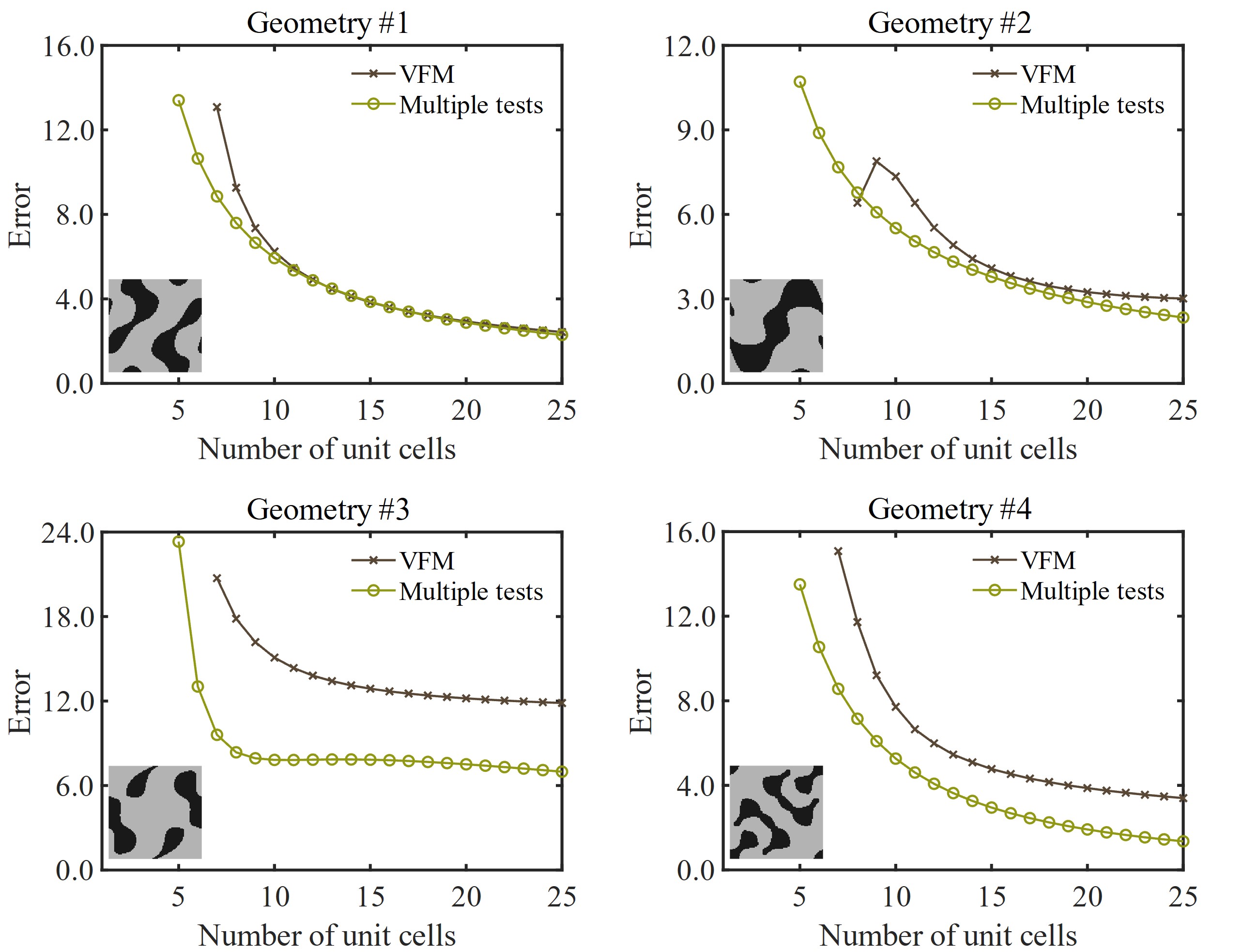}
	\caption{Variation of least square error between homogenized stiffness tensor and stiffness tensor identified using the VFM and the conventional methods as the number of unit cells are varied.}  \label{fig: convergences}
	\end{center} 
\end{figure}

We compare the relative error in parameter identification when performing multiple tests (as in the conventional approach) and when using the VFM (\cref{fig: convergences}).
Since we exclude two rows and columns of boundary unit cells in the proposed VFM, the number of unit cells available to form the system of equations is guaranteed only when there are at least 7 unit cells and the results are shown starting with this number. 
For geometry \#1, as the number of unit cells increases, the error calculated for the conventional method based on multiple tests decreases monotonically from 13.4\% at 5 unit cell {tessellation} to 2.3\% at 25 unit cell {tessellation}. 
Similarly, the error for the VFM decreases monotonically from 13.1\% at 7 unit cell {tessellation} to 2.4\% at 25 unit cell {tessellation}.  
This shows that the parameters identified using our VFM are as good as those obtained by performing multiple tests, as long as there are at least ten repeated unit cells in the domain of interest.
We found this general conclusion to hold for most of the considered geometries.
The only exception is geometry \#3, for which the error remains at 7.0\% (for multiple tests) and 11.8\% (for the VFM) after 10 unit cell {tessellation}. 
In the case of geometry \#3, a major portion of the error lies in just two of the parameters $C_{12}$ and $C_{16}$. These two parameters are quite small relative to the rest of the parameters and hence, they are hard to accurately estimate in comparison to the others.

As the number of unit cells increases, the ratio of the number of unit cells along the boundary to the number of unit cells in the interior decreases.
As a result, the boundary effects described in \cref{sec: synthetic data} diminish and the behavior of the structure approaches the continuum equivalent. 
In \cref{tab: comparion geom1}, we summarize the parameters identified {for geometry \#1} from both the methods against homogenization for 25 unit cell tessellation (see \cref{tab: comparion geom2}, \cref{tab: comparion geom3}, \cref{tab: comparion geom4} for the other geometries).

\begin{table}[!htb]
     \centering
    \begin{tabular}{l|r|r|r|r|r|r} 
    \textbf{Method} & $C_{11}$ (MPa) & $C_{12}$ (MPa)& $C_{22}$ (MPa) & $C_{16}$ (MPa) & $C_{26}$ (MPa) & $C_{66}$ (MPa) \\ \hline
    \text{Homogenization} & 131.62  & 61.98 &  198.38 & 63.58 & 83.87&  95.30\\ 
    \text{VFM} &125.97 & 62.57& 196.15 & 61.35& 82.39& 93.48\\
    \text{Multiple tests} & 129.86 & 67.93& 199.37 & 64.01 &  85.47 & 95.54
    \end{tabular} 
 \caption{Comparison of stiffness tensor parameters identified for geometry \#1 with 25 unit cell tessellation based on synthetic data using the VFM and the conventional methods against the computational homogenization.}
 \label{tab: comparion geom1}
\end{table}

\subsection{Parameter identification based on experimental data}

\begin{figure}[h]
	\begin{center}
		\centering 
	\includegraphics[width=0.8\textwidth]{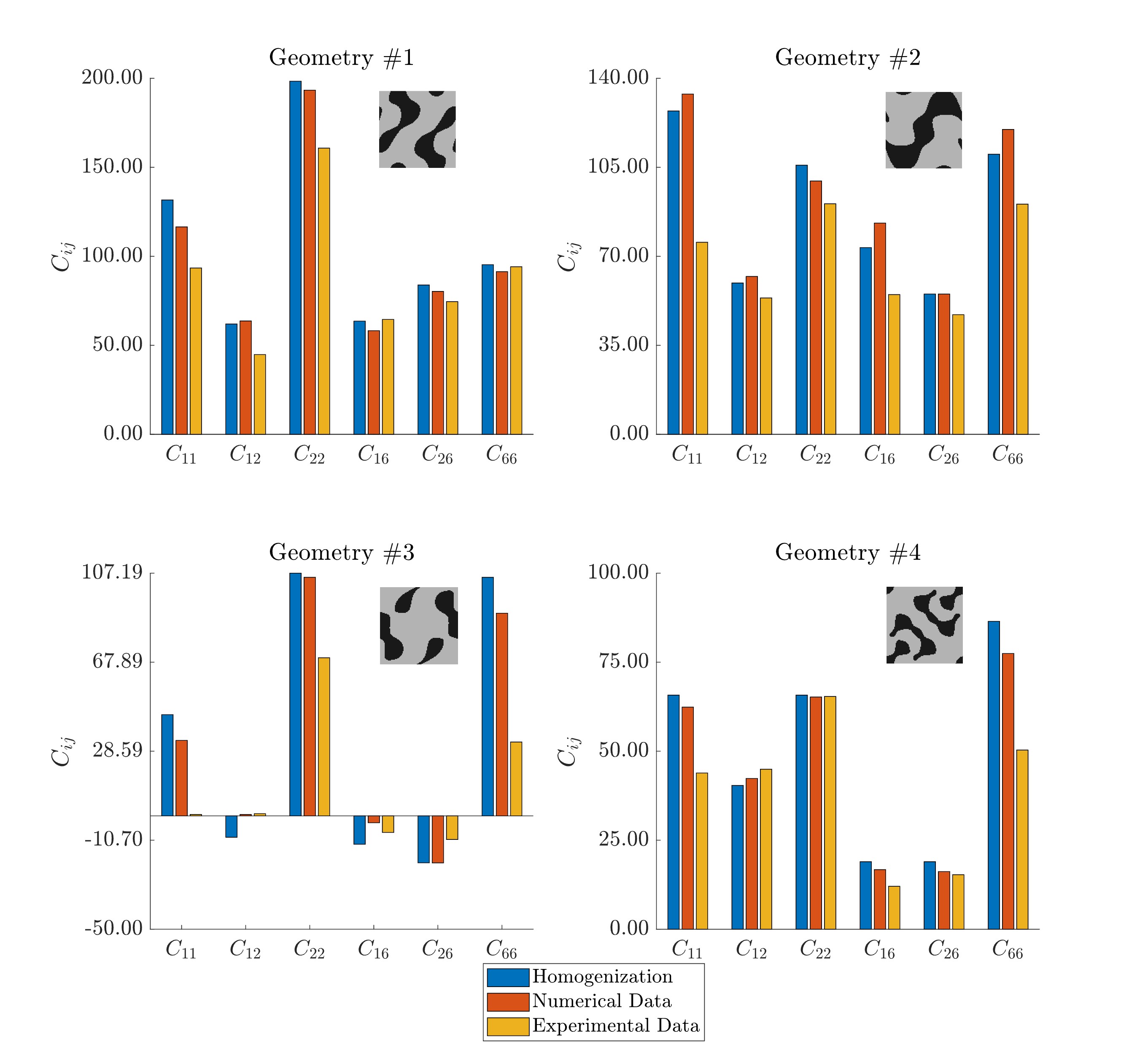}
		\caption{Comparison of material parameters identified using the VFM from numerical and experimental data of 10 unit cell tessellations.}	\label{fig: example geometries results}
	\end{center}
\end{figure}

\cref{fig: example geometries results} summarizes the material parameters identified by the VFM using the simulated and experimental data for 10 unit cell tessellations in comparison to the homogenized stiffness.
The parameters identified using synthetic data compare well with the homogenized properties for all the geometries. 
Further, a good qualitative agreement is observed for the parameters identified using experimental data.
For some of the parameters, such as $C_{12}, C_{22}$, $C_{16}$, $C_{26}$, the experimentally determined parameters match the expectations quantitatively.
In contrast, there is a larger discrepancy in the values of $C_{11}, C_{66}$, for almost all the geometries.
These discrepancies are related to the fact that the experimentally measured displacement $u_1$ appears higher than in the simulations, i.e., about 0.1 mm, leading to an under-prediction of the stiffness in the lateral directions. 
An interesting observation is made for geometry \#4.
Based on the numerical data, we know that $C_{11} = C_{22}$ and $C_{16} = C_{26}$. However, we observe experimentally that $C_{11} < C_{22}$ and $C_{16} < C_{26}$. 
In geometry \#3, the discrepancy may be caused by the architecture itself.
The structure has thin and sharp features in the soft phase. 
As it is well known, the behavior of materials in the vicinity of such sharp discontinuities is quite different from a linear elastic continuum (\cite{rosakis_crack-tip_1986}). 
In fact, the load-displacement data for geometry \#3 show nonlinear behavior (see \cref{fig: loaddata}).
In such micro-structures, our linear elastic model assumption fails. 


\begin{figure}[!htb]
	\begin{center}
		\centering 
		\includegraphics[width=0.8\textwidth]{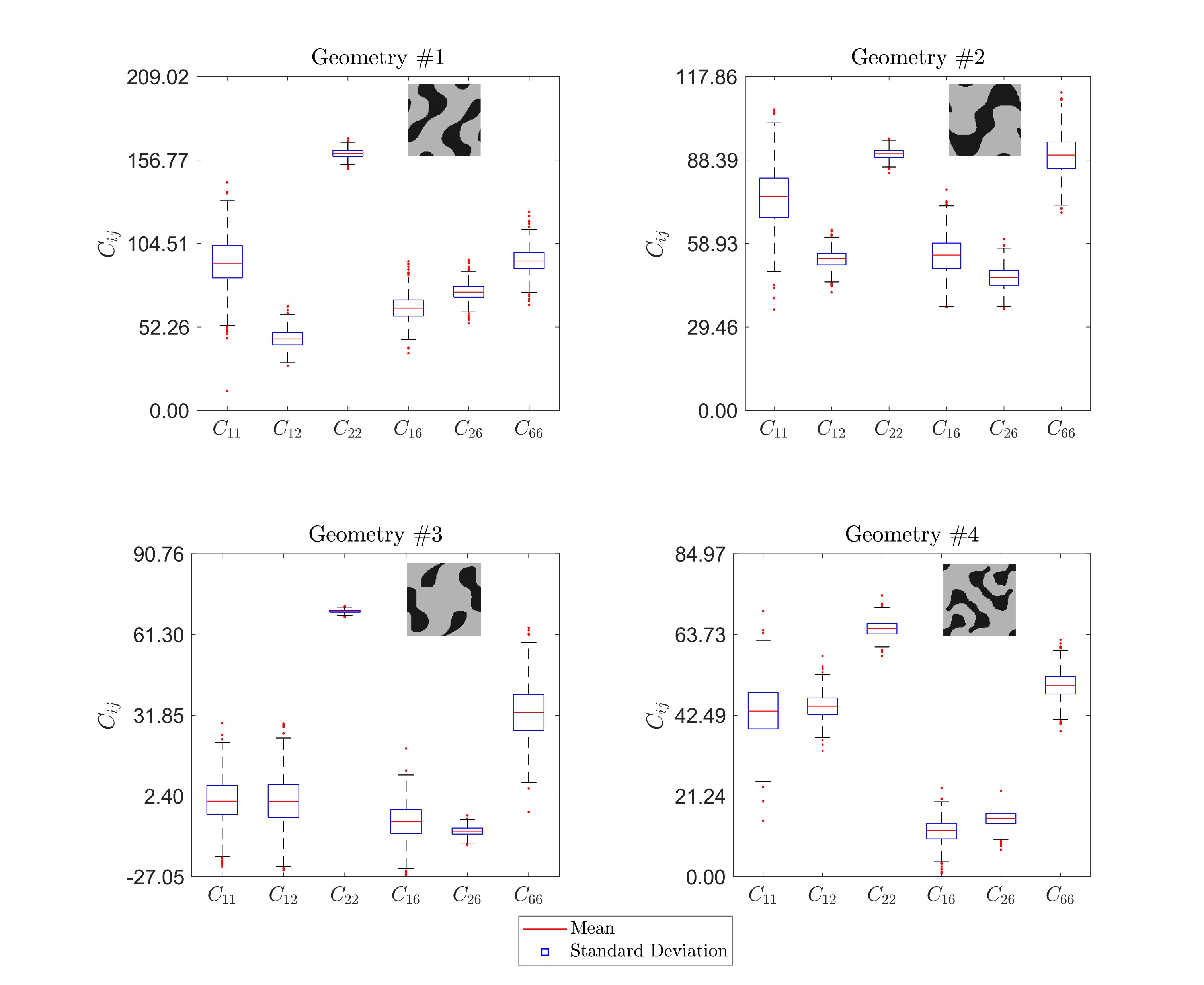}
		\caption{Marginal posterior probability distributions of the material parameters obtained through Bayesian linear regression on the experimental data. The red lines indicate the mean of the marginal posterior distributions. The blue boxes indicate the standard deviation from the mean, i.e., the $68 \%$ probability interval. The black intervals indicate three times the standard deviation from the mean, i.e., the $99 \%$ probability interval.}
		\label{fig:Bayes}
	\end{center}
\end{figure}

We finalize the study by applying the Bayesian method described in \cref{sec:Bayesian_inference} to the experimental data. 
The resulting marginal posterior probability distributions of the material parameters are shown in \cref{fig:Bayes}. 
It is observed that the computed mean values of the marginal posteriors are similar to the deterministic results shown in \cref{fig: example geometries results}. 
Beyond that, the standard deviation of the marginal posteriors indicate (un)certainty in the parameter predictions.
Matching our expectations, the parameter $C_{22}$ is identified with the highest certainty, while for example the identification of the parameter $C_{11}$ shows a high uncertainty. 
Additionally, it is noteworthy that the marginal posteriors of the parameters identified when the Bayesian method is applied to the numerical data show low standard deviations as the data is not affected by the experimental noise (see \cref{fig: BayesFEM}).

We note that, for geometry \#3, the marginal posterior probability distributions of the parameters exceed the thermodynamically admissible range, e.g., the marginal posterior of $C_{11}$ is partially negative. 
This must be considered when interpreting the results.
In this work, no measure was taken to enforce thermodynamic admissibility in the Bayesian method, which thus remains a future objective.

\newpage
\section{Conclusions}
\label{sec: concl}
In this paper, we present an approach to identify the 6 independent elastic material parameters of plane anisotropic elasticity from a single experiment, using the virtual fields method. 
This approach allows identifying shear-normal coupling parameters experimentally, a task that had remained challenging so far.
We first demonstrate the effectiveness of our method using numerically generated data from a single tension test.
We then experimentally validate the method on additively manufactured specimens, by measuring full-field displacement data and traction forces.
We show that our method is effective for materials that include at least 10 repeated unit cells in their structure, to satisfy homogenenization conditions.
We calculate the uncertainity in the identification estimation of the material parameters using Bayesian linear regression.
{In the future, to further refine the experimental parameter identification, it is necessary to optimize the shape of the specimens to ensure strong contributions of strains from different stiffness tensor components.} 
The proposed approach has potential for measurements of elasticity parameters of complex, anisotropic, three-dimensional structured materials and composites with shear-shear couplings, and for the study of their nonlinear behavior.
A further potential application of the method could be for parameter identification of constitutive tensors corresponding to different types of coupled behavior, such as generalized piezoelectric, flexoelectric and piezomagnetic tensors.

\section*{CRediT authorship contribution statement}
\textbf{Jagannadh Boddapati:} Conceptualization, Investigation, Software, Formal analysis, Writing - original Draft. 
\textbf{Moritz Flaschel:} Conceptualization, Investigation, Software, Formal analysis, Writing - original Draft. 
\textbf{Siddhant Kumar:}. Conceptualization, Writing - review and editing.
\textbf{Laura De Lorenzis:} Conceptualization, Writing - review and editing. 
\textbf{Chiara Daraio:}. Conceptualization, Writing - review and editing, Supervision, Funding acquisition.
\section*{Acknowledgements}
{We thank Pierre Margerit (École Polytechnique) for the discussion on digital image correlation, and
Jihoon Ahn (Caltech) and Perry Samimy (Caltech) for their help on designing the experiments.
C.D. and J.B. acknowledge support from the US National Science Foundation (NSF), grant number 1835735.
M.F. and L.D.L. acknowledge support from the Swiss National Science Foundation (SNF), project number 200021\_204316.}

\newpage
\appendix

\newpage
\section{Parameter identification based on multiple tests}
\label{sec: multiple tests}

\begin{figure}[!htb]
	\begin{center}
		\centering 
		\includegraphics[width=0.9\textwidth]{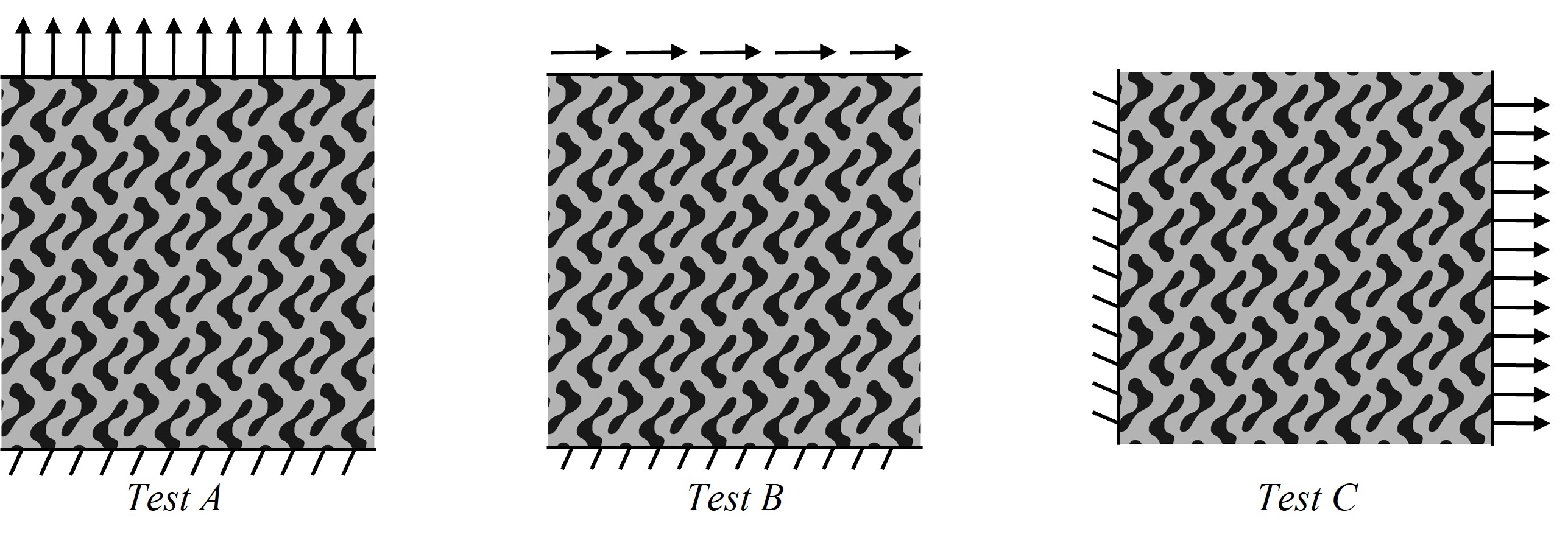}
		\caption{Parameter identification of an anisotropic metamaterial by performing three different tests.}	\label{fig: three boundary conditions}
	\end{center}
\end{figure}

In this section, we explore a method of parameter identification that involves multiple tests (as in the conventional approach) in the context of anisotropic metamaterials.
We subject the metamaterial to three different tests namely \textit{Test  A},  \textit{Test  B}, and \textit{Test  C} as shown in \cref{fig: three boundary conditions}. 
\textit{Test A} and \textit{Test  C} are tension tests along $x_2$ and $x_1$ axis respectively, and \textit{Test B} is a simple shear test.
We assume that the average strains $\tilde{\bm{\varepsilon}}_{ij}^{A,B,C}$ are known experimentally from full-field measurements.
In addition, the reaction forces at the fixed end are known experimentally from load sensor measurements. 
We will show that the material parameters can be identified from the average strains and the net reaction forces from these three tests. 

From Gauss' divergence theorem, the average stresses $\tilde{\bm{\sigma}}$ are related to the tractions $\bm{t}$ at the fixed end as
\begin{equation}
    t_i = \tilde{\sigma}_{ij}n_j, \label{eq: stress to traction}
\end{equation}
where $\bm{n}$ is the unit outward normal. 
For \textit{Test A}, the unit outward normal $\boldsymbol{n}$ at the fixed end is $[0,-1]^T$. 
Using \cref{eq: stress to traction,eq: ConstIndicialSmall}, and assuming homogenized effective continuum behavior for the structured solid, we get 
\begin{subequations}
\begin{align}
F_{1}^{A} /\mathcal{A} &= \tilde{\sigma}_{6}^{A}=C_{16} \tilde{\varepsilon}_{11}^{A}+C_{26} \tilde{\varepsilon}_{2}^{A}+C_{66}\left(2 \tilde{\varepsilon}_{6}^{A}\right) \label{eq: Fx Test1},\\
F_{2}^{A} /\mathcal{A} &=\tilde{\sigma}_{2}^{A}=C_{12} \tilde{\varepsilon}_{1}^{A}+C_{22} \tilde{\varepsilon}_{2}^{A}+C_{26}\left(2\tilde{\varepsilon}_{6}^{A}\right), \label{eq: Fy Test1}
\end{align}
\end{subequations}
where $\tilde{\sigma}_{12}^A,\tilde{\sigma}_{22}^A$ are the average stress components, $F_1^A,F_2^A$ are the reaction force components at the fixed end from \textit{Test A} and $\mathcal{A}$ is the cross sectional area of the fixed end. 

Similarly, from  \textit{Test B} and \textit{Test C}, we get 
\begin{subequations}
\begin{align}
F_{1}^{B} /\mathcal{A}&=\tilde{\sigma}_{6}^{B}=C_{16} \tilde{\varepsilon}_{1}^{B}+C_{26} \tilde{\varepsilon}_{2}^{B}+C_{66}\left(2 \tilde{\varepsilon}_{6}^{B}\right) \label{eq: Fx Test2},\\
F_{2}^{B} /\mathcal{A}&=\tilde{\sigma}_{2}^{B}=C_{12} \tilde{\varepsilon}_{1}^{B}+C_{22} \tilde{\varepsilon}_{2}^{B}+C_{26}\left(2 \tilde{\varepsilon}_{6}^{B}\right) \label{eq: Fy Test2},\\
F_{1}^{C} /\mathcal{A}&=\tilde{\sigma}_{1}^{C}=C_{11} \tilde{\varepsilon}_{1}^{C}+C_{12} \tilde{\varepsilon}_{2}^{C}+C_{16}\left(2 \tilde{\varepsilon}_{6}^{C}\right) \label{eq: Fx Test3},\\
F_{2}^{C} /\mathcal{A}&=\tilde{\sigma}_{6}^{C}=C_{16} \tilde{\varepsilon}_{1}^{C}+C_{26} \tilde{\varepsilon}_{2}^{C}+C_{66}\left(2 \tilde{\varepsilon}_{6}^{C}\right) \label{eq: Fy Test3}.
\end{align}
\end{subequations}
Rearranging \cref{eq: Fx Test1,eq: Fy Test1,eq: Fx Test2,eq: Fy Test2,eq: Fx Test3,eq: Fy Test3} into a matrix form, we obtain a system of linear equations, 
\begin{equation}
\left[\begin{array}{cccccc}
0 & 0 & 2 \tilde{\varepsilon}_{6}^{A} & \tilde{\varepsilon}_{1}^{A} & \tilde{\varepsilon}_{2}^{A} & 0 \\
0 & \tilde{\varepsilon}_{2}^{A} & 0 & 0 & 2 \tilde{\varepsilon}_{6}^{A} & \tilde{\varepsilon}_{1}^{A} \\
0 & 0 & 2 \tilde{\varepsilon}_{6}^{B} & \tilde{\varepsilon}_{1}^{B} & \tilde{\varepsilon}_{2}^{B} & 0 \\
0 & \tilde{\varepsilon}_{2}^{B} & 0 & 0 & 2 \tilde{\varepsilon}_{6}^{B} & \tilde{\varepsilon}_{1}^{B} \\
\tilde{\varepsilon}_{1}^{C} & 0 & 0 & 2 \tilde{\varepsilon}_{6}^{C} & 0 & \tilde{\varepsilon}_{2}^{C} \\
0 & 0 & 2 \tilde{\varepsilon}_{6}^{C} & \tilde{\varepsilon}_{1}^{C} & \tilde{\varepsilon}_{2} & 0
\end{array}\right]\left[\begin{array}{l}
C_{11} \\
C_{22} \\
C_{66} \\
C_{16} \\
C_{26} \\
C_{12}
\end{array}\right] = \frac{1}{\mathcal{A}}\left[\begin{array}{c}
F_{1}^{A} \\
F_{2}^{A} \\
F_{1}^{B} \\
F_{2}^{B} \\
F_{1}^{C} \\
F_{2}^{C}
\end{array}\right].\label{eq: system 3BC}
\end{equation} 

For readability, \cref{eq: system 3BC} is written as 
\begin{equation}
    \mathcal{A}\tilde{\bm{\varepsilon}}\bm{C}_\text{vec} = \bm{F}_\text{vec} ,\label{eq: system 3BC summ}
\end{equation}
where $\tilde{\bm{\varepsilon}}$ is a non-symmetric square matrix of size 6 containing average strain components from all of the tests and $\bm{F}_\text{vec}$ is a vector containing net reaction force components from all of the tests. 
Then the material parameters $\bm{C}^{\text{opt}}_{\text{vec}}$ can be obtained as a solution to the least squares minimization problem, 
\begin{equation}
 \bm{C}^{\text{opt}}_{\text{vec}} = \argminWithArgs_{\bm{C}_{\text{vec}}} {\|\mathcal{A}\tilde{\bm{\varepsilon}}\bm{C}_\text{{vec}} -\bm{F}_\text{vec}\|^2} .\label{eq: system 3BC summ solution}
\end{equation}

 It should be noted that we use this method for material parameter identification, only using the numerical data.
 We did not experimentally validate this method, since shear testing is non-trivial and requires dedicated setups, such as a hexapod machine \citep{dalemat}.


\section{Additional data}
\begin{table}[!htb]
    \centering
    \begin{tabular}{l|r|r|r|r|r|r} 
    \textbf{Method} & $C_{11}$ (MPa) & $C_{12}$ (MPa)& $C_{22}$ (MPa) & $C_{16}$ (MPa) & $C_{26}$ (MPa) & $C_{66}$ (MPa) \\ \hline
    \text{Homogenization} & 127.14&59.50&105.83&73.42&55.16&110.15\\
    \text{VFM} &129.92&58.84&102.72&77.43&55.25&113.73\\ 
    \text{Multiple tests} & 123.82&57.44&102.35&72.74&54.68&110.12
    \end{tabular} 
     \caption{Comparison of stiffness tensor parameters identified for geometry \#2 with 25 unit cell tessellation based on synthetic data using the VFM and the conventional methods against the computational homogenization.}
 \label{tab: comparion geom2}
\end{table}

\begin{table}[!htb]
     \centering
    \begin{tabular}{l|r|r|r|r|r|r} 
    \textbf{Method} & $C_{11}$ (MPa) & $C_{12}$ (MPa)& $C_{22}$ (MPa) & $C_{16}$ (MPa) & $C_{26}$ (MPa) & $C_{66}$ (MPa) \\ \hline
    \text{Homogenization} & 44.70&-9.42&107.19&-12.52&-20.71&105.35\\ 
    \text{VFM} &33.95&-5.84&106.48&-4.44&-20.78&92.65\\
    \text{Multiple tests} & 45.78&-19.27&112.01&-11.30&-20.30&105.65
    \end{tabular} 
    \caption{Comparison of stiffness tensor parameters identified for geometry \#3 with 25 unit cell tessellation based on synthetic data using the VFM and the conventional methods against the computational homogenization.}
 \label{tab: comparion geom3}
\end{table}

\begin{table}[!htb]
    \centering
    \begin{tabular}{l|r|r|r|r|r|r} 
    \textbf{Method} & $C_{11}$ (MPa) & $C_{12}$ (MPa)& $C_{22}$ (MPa) & $C_{16}$ (MPa) & $C_{26}$ (MPa) & $C_{66}$ (MPa) \\ \hline
    \text{Homogenization} & 65.74&40.36&65.74&18.95&18.95&86.47\\
    \text{VFM} &65.45&40.94&65.43&17.93&17.76&82.19\\
    \text{Multiple tests} & 66.66&41.58&66.66&18.75&18.75&86.86
    \end{tabular} 
    \caption{Comparison of stiffness tensor parameters identified for geometry \#4 with 25 unit cell tessellation based on synthetic data using the VFM and the conventional methods against the computational homogenization.}
 \label{tab: comparion geom4}
\end{table}

\newpage
\begin{figure}[!htb]
	\begin{center}
		\centering 
	\includegraphics[width=0.75\textwidth]{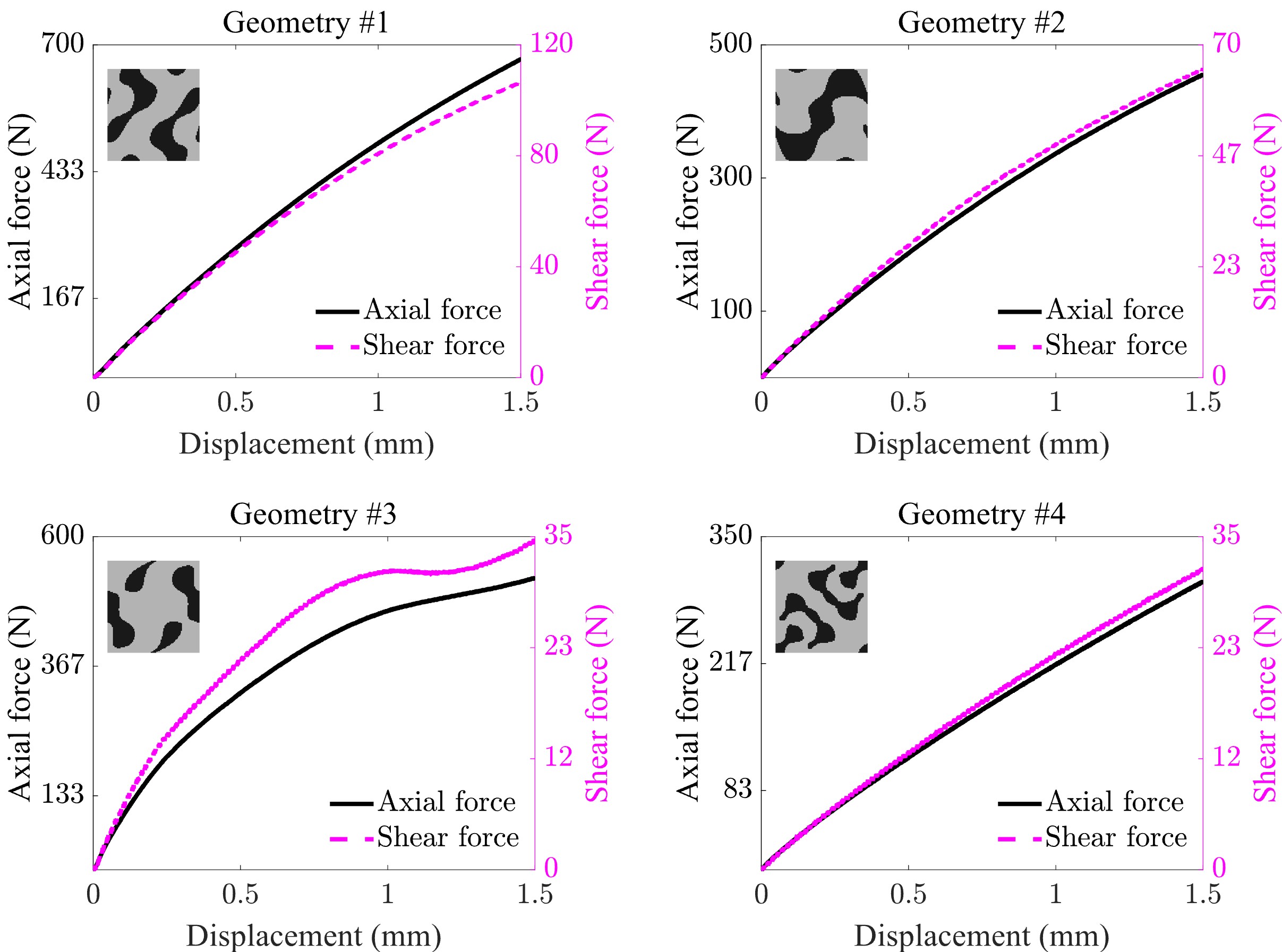}
		\caption{Axial and shear load-displacement data for all the experimentally tested specimens.}
		\label{fig: loaddata}
	\end{center}
\end{figure}

\begin{figure}[!htb]
	\begin{center}
		\centering 
		\includegraphics[width=0.8\textwidth]{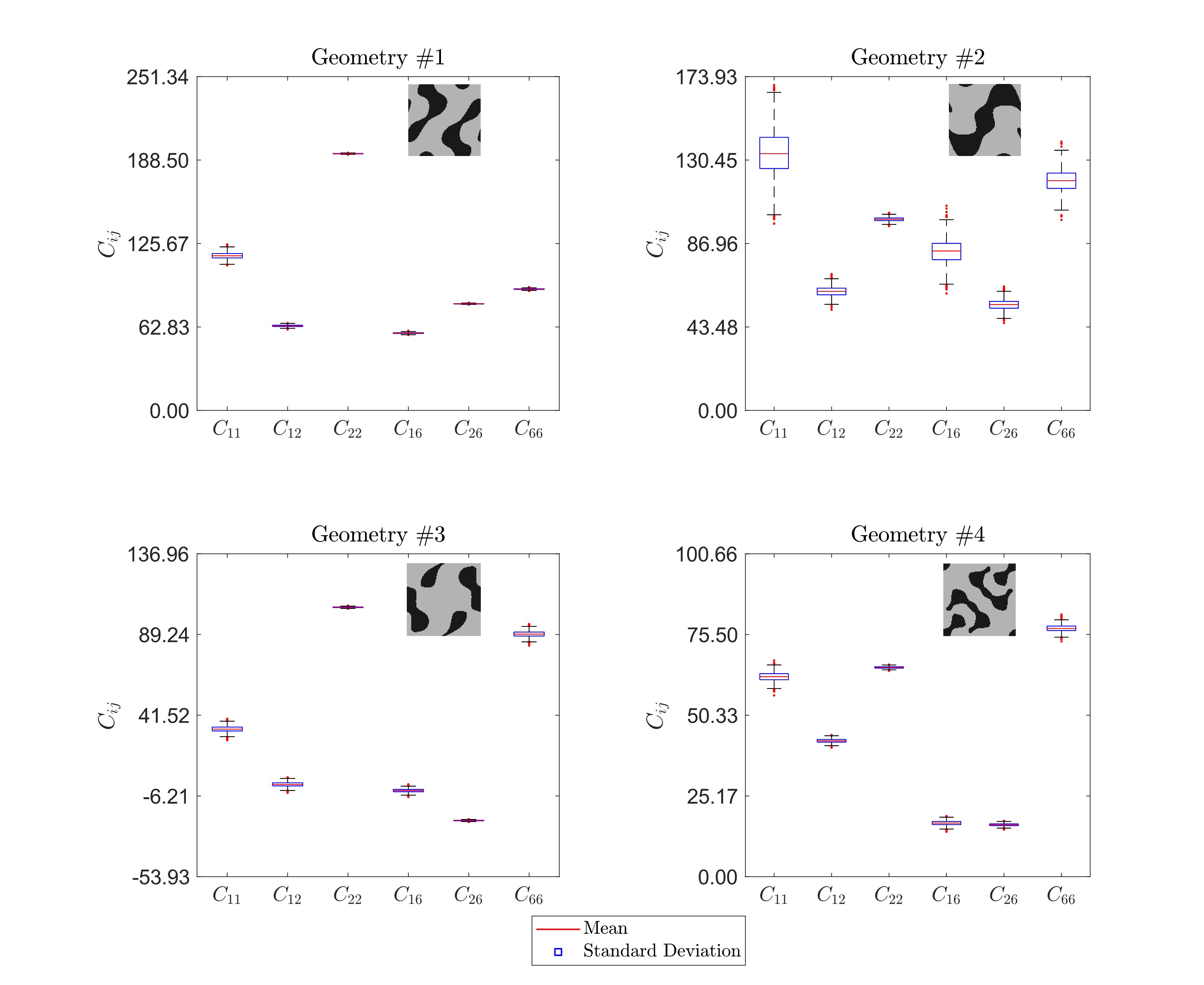}
		\caption{Marginal posterior probability distributions of the material parameters obtained through Bayesian linear regression on the numerical data.}
		\label{fig: BayesFEM}
	\end{center}
\end{figure}

\newpage
\bibliographystyle{elsarticle-harv}
\bibliography{BibJB, BibMF}

\begin{thebibliography}{73}
\expandafter\ifx\csname natexlab\endcsname\relax\def\natexlab#1{#1}\fi
\providecommand{\url}[1]{\texttt{#1}}
\providecommand{\href}[2]{#2}
\providecommand{\path}[1]{#1}
\providecommand{\DOIprefix}{doi:}
\providecommand{\ArXivprefix}{arXiv:}
\providecommand{\URLprefix}{URL: }
\providecommand{\Pubmedprefix}{pmid:}
\providecommand{\doi}[1]{\href{http://dx.doi.org/#1}{\path{#1}}}
\providecommand{\Pubmed}[1]{\href{pmid:#1}{\path{#1}}}
\providecommand{\bibinfo}[2]{#2}
\ifx\xfnm\relax \def\xfnm[#1]{\unskip,\space#1}\fi
\bibitem[{Agnelli et~al.(2021)Agnelli, Margerit, Celli, Daraio and
  Constantinescu}]{agnelli_systematic_2021}
\bibinfo{author}{Agnelli, F.}, \bibinfo{author}{Margerit, P.},
  \bibinfo{author}{Celli, P.}, \bibinfo{author}{Daraio, C.},
  \bibinfo{author}{Constantinescu, A.}, \bibinfo{year}{2021}.
\newblock \bibinfo{title}{Systematic two-scale image analysis of extreme
  deformations in soft architectured sheets}.
\newblock \bibinfo{journal}{International Journal of Mechanical Sciences}
  \bibinfo{volume}{194}, \bibinfo{pages}{106205}.
\newblock \URLprefix
  \url{https://www.sciencedirect.com/science/article/pii/S0020740320343101},
  \DOIprefix\doi{10.1016/j.ijmecsci.2020.106205}.
\bibitem[{Agnelli et~al.(2022)Agnelli, Nika and
  Constantinescu}]{agnelli_design_2022}
\bibinfo{author}{Agnelli, F.}, \bibinfo{author}{Nika, G.},
  \bibinfo{author}{Constantinescu, A.}, \bibinfo{year}{2022}.
\newblock \bibinfo{title}{Design of thin micro-architectured panels with
  extension–bending coupling effects using topology optimization}.
\newblock \bibinfo{journal}{Computer Methods in Applied Mechanics and
  Engineering} \bibinfo{volume}{391}, \bibinfo{pages}{114496}.
\newblock \URLprefix
  \url{https://www.sciencedirect.com/science/article/pii/S0045782521007040},
  \DOIprefix\doi{10.1016/j.cma.2021.114496}.
\bibitem[{Ahn et~al.(2017)Ahn, Lee and Kim}]{ahn_conical_2017}
\bibinfo{author}{Ahn, Y.K.}, \bibinfo{author}{Lee, H.J.}, \bibinfo{author}{Kim,
  Y.Y.}, \bibinfo{year}{2017}.
\newblock \bibinfo{title}{Conical {Refraction} of {Elastic} {Waves} by
  {Anisotropic} {Metamaterials} and {Application} for {Parallel} {Translation}
  of {Elastic} {Waves}}.
\newblock \bibinfo{journal}{Scientific Reports} \bibinfo{volume}{7},
  \bibinfo{pages}{10072}.
\newblock \URLprefix \url{http://www.nature.com/articles/s41598-017-10691-6},
  \DOIprefix\doi{10.1038/s41598-017-10691-6}.
\bibitem[{Andreassen and Andreasen(2014)}]{andreassen_how_2014}
\bibinfo{author}{Andreassen, E.}, \bibinfo{author}{Andreasen, C.S.},
  \bibinfo{year}{2014}.
\newblock \bibinfo{title}{How to determine composite material properties using
  numerical homogenization}.
\newblock \bibinfo{journal}{Computational Materials Science}
  \bibinfo{volume}{83}, \bibinfo{pages}{488--495}.
\newblock \URLprefix
  \url{https://www.sciencedirect.com/science/article/pii/S0927025613005302},
  \DOIprefix\doi{10.1016/j.commatsci.2013.09.006}.
\bibitem[{Andreaus et~al.(2016)Andreaus, dell’Isola, Giorgio, Placidi,
  Lekszycki and Rizzi}]{andreaus_numerical_2016}
\bibinfo{author}{Andreaus, U.}, \bibinfo{author}{dell’Isola, F.},
  \bibinfo{author}{Giorgio, I.}, \bibinfo{author}{Placidi, L.},
  \bibinfo{author}{Lekszycki, T.}, \bibinfo{author}{Rizzi, N.L.},
  \bibinfo{year}{2016}.
\newblock \bibinfo{title}{Numerical simulations of classical problems in
  two-dimensional (non) linear second gradient elasticity}.
\newblock \bibinfo{journal}{International Journal of Engineering Science}
  \bibinfo{volume}{108}, \bibinfo{pages}{34--50}.
\newblock \URLprefix
  \url{https://linkinghub.elsevier.com/retrieve/pii/S0020722516308187},
  \DOIprefix\doi{10.1016/j.ijengsci.2016.08.003}.
\bibitem[{Auffray and Ropars(2016)}]{auffray_invariant-based_2016}
\bibinfo{author}{Auffray, N.}, \bibinfo{author}{Ropars, P.},
  \bibinfo{year}{2016}.
\newblock \bibinfo{title}{Invariant-based reconstruction of bidimensional
  elasticity tensors}.
\newblock \bibinfo{journal}{International Journal of Solids and Structures}
  \bibinfo{volume}{87}, \bibinfo{pages}{183--193}.
\newblock \URLprefix
  \url{https://linkinghub.elsevier.com/retrieve/pii/S0020768316000676},
  \DOIprefix\doi{10.1016/j.ijsolstr.2016.02.013}.
\bibitem[{Avril et~al.(2008)Avril, Bonnet, Bretelle, Grédiac, Hild, Ienny,
  Latourte, Lemosse, Pagano, Pagnacco and Pierron}]{avril_overview_2008}
\bibinfo{author}{Avril, S.}, \bibinfo{author}{Bonnet, M.},
  \bibinfo{author}{Bretelle, A.S.}, \bibinfo{author}{Grédiac, M.},
  \bibinfo{author}{Hild, F.}, \bibinfo{author}{Ienny, P.},
  \bibinfo{author}{Latourte, F.}, \bibinfo{author}{Lemosse, D.},
  \bibinfo{author}{Pagano, S.}, \bibinfo{author}{Pagnacco, E.},
  \bibinfo{author}{Pierron, F.}, \bibinfo{year}{2008}.
\newblock \bibinfo{title}{Overview of {Identification} {Methods} of
  {Mechanical} {Parameters} {Based} on {Full}-field {Measurements}}.
\newblock \bibinfo{journal}{Experimental Mechanics} \bibinfo{volume}{48},
  \bibinfo{pages}{381--402}.
\newblock \URLprefix \url{http://link.springer.com/10.1007/s11340-008-9148-y},
  \DOIprefix\doi{10.1007/s11340-008-9148-y}.
\bibitem[{Avril et~al.(2004)Avril, Grédiac and
  Pierron}]{avril_sensitivity_2004}
\bibinfo{author}{Avril, S.}, \bibinfo{author}{Grédiac, M.},
  \bibinfo{author}{Pierron, F.}, \bibinfo{year}{2004}.
\newblock \bibinfo{title}{Sensitivity of the virtual fields method to noisy
  data}.
\newblock \bibinfo{journal}{Computational Mechanics} \bibinfo{volume}{34},
  \bibinfo{pages}{439--452}.
\newblock \URLprefix \url{https://doi.org/10.1007/s00466-004-0589-6},
  \DOIprefix\doi{10.1007/s00466-004-0589-6}.
\bibitem[{Bastek et~al.(2022)Bastek, Kumar, Telgen, Glaesener and
  Kochmann}]{bastek_inverting_2022}
\bibinfo{author}{Bastek, J.H.}, \bibinfo{author}{Kumar, S.},
  \bibinfo{author}{Telgen, B.}, \bibinfo{author}{Glaesener, R.N.},
  \bibinfo{author}{Kochmann, D.M.}, \bibinfo{year}{2022}.
\newblock \bibinfo{title}{Inverting the structure–property map of truss
  metamaterials by deep learning}.
\newblock \bibinfo{journal}{Proceedings of the National Academy of Sciences}
  \bibinfo{volume}{119}, \bibinfo{pages}{e2111505119}.
\newblock \URLprefix
  \url{https://www.pnas.org/doi/abs/10.1073/pnas.2111505119},
  \DOIprefix\doi{10.1073/pnas.2111505119}. \bibinfo{note}{publisher:
  Proceedings of the National Academy of Sciences}.
\bibitem[{Bertoldi et~al.(2017)Bertoldi, Vitelli, Christensen and van
  Hecke}]{bertoldi_flexible_2017}
\bibinfo{author}{Bertoldi, K.}, \bibinfo{author}{Vitelli, V.},
  \bibinfo{author}{Christensen, J.}, \bibinfo{author}{van Hecke, M.},
  \bibinfo{year}{2017}.
\newblock \bibinfo{title}{Flexible mechanical metamaterials}.
\newblock \bibinfo{journal}{Nature Reviews Materials} \bibinfo{volume}{2},
  \bibinfo{pages}{1--11}.
\newblock \URLprefix \url{https://www.nature.com/articles/natrevmats201766},
  \DOIprefix\doi{10.1038/natrevmats.2017.66}. \bibinfo{note}{number: 11
  Publisher: Nature Publishing Group}.
\bibitem[{Besnard et~al.(2006)Besnard, Hild and
  Roux}]{besnard_finite-element_2006}
\bibinfo{author}{Besnard, G.}, \bibinfo{author}{Hild, F.},
  \bibinfo{author}{Roux, S.}, \bibinfo{year}{2006}.
\newblock \bibinfo{title}{“{Finite}-{Element}” {Displacement} {Fields}
  {Analysis} from {Digital} {Images}: {Application} to {Portevin}–{Le}
  {Châtelier} {Bands}}.
\newblock \bibinfo{journal}{Experimental Mechanics} \bibinfo{volume}{46},
  \bibinfo{pages}{789--803}.
\newblock \URLprefix \url{https://doi.org/10.1007/s11340-006-9824-8},
  \DOIprefix\doi{10.1007/s11340-006-9824-8}.
\bibitem[{Cahn(1965)}]{cahn_phase_1965}
\bibinfo{author}{Cahn, J.W.}, \bibinfo{year}{1965}.
\newblock \bibinfo{title}{Phase {Separation} by {Spinodal} {Decomposition} in
  {Isotropic} {Systems}}.
\newblock \bibinfo{journal}{The Journal of Chemical Physics}
  \bibinfo{volume}{42}, \bibinfo{pages}{93--99}.
\newblock \URLprefix \url{https://aip.scitation.org/doi/10.1063/1.1695731},
  \DOIprefix\doi{10.1063/1.1695731}. \bibinfo{note}{publisher: American
  Institute of Physics}.
\bibitem[{Chen et~al.(2018)Chen, Ruan and Huang}]{chen_optimization_2018}
\bibinfo{author}{Chen, W.}, \bibinfo{author}{Ruan, D.}, \bibinfo{author}{Huang,
  X.}, \bibinfo{year}{2018}.
\newblock \bibinfo{title}{Optimization for twist chirality of structural
  materials induced by axial strain}.
\newblock \bibinfo{journal}{Materials Today Communications}
  \bibinfo{volume}{15}, \bibinfo{pages}{175--184}.
\newblock \URLprefix
  \url{https://www.sciencedirect.com/science/article/pii/S235249281830031X},
  \DOIprefix\doi{10.1016/j.mtcomm.2018.03.010}.
\bibitem[{Christensen et~al.(2015)Christensen, Kadic, Wegener, Kraft and
  Wegener}]{christensen_vibrant_2015}
\bibinfo{author}{Christensen, J.}, \bibinfo{author}{Kadic, M.},
  \bibinfo{author}{Wegener, M.}, \bibinfo{author}{Kraft, O.},
  \bibinfo{author}{Wegener, M.}, \bibinfo{year}{2015}.
\newblock \bibinfo{title}{Vibrant times for mechanical metamaterials}.
\newblock \bibinfo{journal}{MRS Communications} \bibinfo{volume}{5},
  \bibinfo{pages}{453--462}.
\newblock \URLprefix \url{http://link.springer.com/10.1557/mrc.2015.51},
  \DOIprefix\doi{10.1557/mrc.2015.51}.
\bibitem[{Dalemat(2019)}]{dalemat}
\bibinfo{author}{Dalemat, M.}, \bibinfo{year}{2019}.
\newblock \bibinfo{title}{{Une exp{\'e}rimentation r{\'e}ussie pour
  l'identification de la r{\'e}ponse m{\'e}canique sans loi de comportement :
  Approche data-driven appliqu{\'e}e aux membranes {\'e}lastom{\`e}res}}.
\newblock \bibinfo{type}{Theses}. {{\'E}cole centrale de Nantes}.
\newblock \URLprefix \url{https://tel.archives-ouvertes.fr/tel-02506891}.
\bibitem[{Dos~Reis and Karathanasopoulos(2022)}]{dos_reis_inverse_2022}
\bibinfo{author}{Dos~Reis, F.}, \bibinfo{author}{Karathanasopoulos, N.},
  \bibinfo{year}{2022}.
\newblock \bibinfo{title}{Inverse metamaterial design combining genetic
  algorithms with asymptotic homogenization schemes}.
\newblock \bibinfo{journal}{International Journal of Solids and Structures}
  \bibinfo{volume}{250}, \bibinfo{pages}{111702}.
\newblock \URLprefix
  \url{https://www.sciencedirect.com/science/article/pii/S0020768322002104},
  \DOIprefix\doi{10.1016/j.ijsolstr.2022.111702}.
\bibitem[{Every and Sachse(1990)}]{every_determination_1990}
\bibinfo{author}{Every, A.G.}, \bibinfo{author}{Sachse, W.},
  \bibinfo{year}{1990}.
\newblock \bibinfo{title}{Determination of the elastic constants of anisotropic
  solids from acoustic-wave group-velocity measurements}.
\newblock \bibinfo{journal}{Physical Review B} \bibinfo{volume}{42},
  \bibinfo{pages}{8196--8205}.
\newblock \URLprefix \url{https://link.aps.org/doi/10.1103/PhysRevB.42.8196},
  \DOIprefix\doi{10.1103/PhysRevB.42.8196}. \bibinfo{note}{publisher: American
  Physical Society}.
\bibitem[{Fischer et~al.(2011)Fischer, Klassen, Mergheim, Steinmann and
  Müller}]{fischer_isogeometric_2011}
\bibinfo{author}{Fischer, P.}, \bibinfo{author}{Klassen, M.},
  \bibinfo{author}{Mergheim, J.}, \bibinfo{author}{Steinmann, P.},
  \bibinfo{author}{Müller, R.}, \bibinfo{year}{2011}.
\newblock \bibinfo{title}{Isogeometric analysis of {2D} gradient elasticity}.
\newblock \bibinfo{journal}{Computational Mechanics} \bibinfo{volume}{47},
  \bibinfo{pages}{325--334}.
\newblock \URLprefix \url{http://link.springer.com/10.1007/s00466-010-0543-8},
  \DOIprefix\doi{10.1007/s00466-010-0543-8}.
\bibitem[{Flaschel(2023)}]{flaschel_automated_2023-1}
\bibinfo{author}{Flaschel, M.}, \bibinfo{year}{2023}.
\newblock \bibinfo{title}{Automated {Discovery} of {Material} {Models} in
  {Continuum} {Solid} {Mechanics}}.
\newblock Ph.D. thesis. ETH Zurich.
\newblock \URLprefix \url{http://hdl.handle.net/20.500.11850/602750},
  \DOIprefix\doi{10.3929/ETHZ-B-000602750}.
\bibitem[{Flaschel et~al.(2021)Flaschel, Kumar and
  De~Lorenzis}]{flaschel_unsupervised_2021}
\bibinfo{author}{Flaschel, M.}, \bibinfo{author}{Kumar, S.},
  \bibinfo{author}{De~Lorenzis, L.}, \bibinfo{year}{2021}.
\newblock \bibinfo{title}{Unsupervised discovery of interpretable hyperelastic
  constitutive laws}.
\newblock \bibinfo{journal}{Computer Methods in Applied Mechanics and
  Engineering} \bibinfo{volume}{381}, \bibinfo{pages}{113852}.
\newblock \DOIprefix\doi{10.1016/j.cma.2021.113852}.
\bibitem[{Flaschel et~al.(2022)Flaschel, Kumar and
  De~Lorenzis}]{flaschel_discovering_2022}
\bibinfo{author}{Flaschel, M.}, \bibinfo{author}{Kumar, S.},
  \bibinfo{author}{De~Lorenzis, L.}, \bibinfo{year}{2022}.
\newblock \bibinfo{title}{Discovering plasticity models without stress data}.
\newblock \bibinfo{journal}{npj Computational Materials} \bibinfo{volume}{8},
  \bibinfo{pages}{91}.
\newblock \URLprefix \url{https://www.nature.com/articles/s41524-022-00752-4},
  \DOIprefix\doi{10.1038/s41524-022-00752-4}.
\bibitem[{Flaschel et~al.(2023)Flaschel, Kumar and
  De~Lorenzis}]{flaschel_automated_2023}
\bibinfo{author}{Flaschel, M.}, \bibinfo{author}{Kumar, S.},
  \bibinfo{author}{De~Lorenzis, L.}, \bibinfo{year}{2023}.
\newblock \bibinfo{title}{Automated discovery of generalized standard material
  models with {EUCLID}}.
\newblock \bibinfo{journal}{Computer Methods in Applied Mechanics and
  Engineering} \bibinfo{volume}{405}, \bibinfo{pages}{115867}.
\newblock \URLprefix
  \url{https://linkinghub.elsevier.com/retrieve/pii/S0045782522008234},
  \DOIprefix\doi{10.1016/j.cma.2022.115867}.
\bibitem[{Forte and Vianello(2014)}]{forte_unified_2014}
\bibinfo{author}{Forte, S.}, \bibinfo{author}{Vianello, M.},
  \bibinfo{year}{2014}.
\newblock \bibinfo{title}{A unified approach to invariants of plane elasticity
  tensors}.
\newblock \bibinfo{journal}{Meccanica} \bibinfo{volume}{49},
  \bibinfo{pages}{2001--2012}.
\newblock \URLprefix \url{http://link.springer.com/10.1007/s11012-014-9916-y},
  \DOIprefix\doi{10.1007/s11012-014-9916-y}.
\bibitem[{François et~al.(1998)François, Geymonat and
  Berthaud}]{francois_determination_1998}
\bibinfo{author}{François, M.}, \bibinfo{author}{Geymonat, G.},
  \bibinfo{author}{Berthaud, Y.}, \bibinfo{year}{1998}.
\newblock \bibinfo{title}{Determination of the symmetries of an experimentally
  determined stiffness tensor: {Application} to acoustic measurements}.
\newblock \bibinfo{journal}{International Journal of Solids and Structures}
  \bibinfo{volume}{35}, \bibinfo{pages}{4091--4106}.
\newblock \URLprefix
  \url{https://www.sciencedirect.com/science/article/pii/S002076839700303X},
  \DOIprefix\doi{10.1016/S0020-7683(97)00303-X}.
\bibitem[{Frenzel et~al.(2017)Frenzel, Kadic and
  Wegener}]{frenzel_three-dimensional_2017}
\bibinfo{author}{Frenzel, T.}, \bibinfo{author}{Kadic, M.},
  \bibinfo{author}{Wegener, M.}, \bibinfo{year}{2017}.
\newblock \bibinfo{title}{Three-dimensional mechanical metamaterials with a
  twist}.
\newblock \bibinfo{journal}{Science} \bibinfo{volume}{358},
  \bibinfo{pages}{1072--1074}.
\newblock \URLprefix
  \url{https://www.sciencemag.org/lookup/doi/10.1126/science.aao4640},
  \DOIprefix\doi{10.1126/science.aao4640}.
\bibitem[{Gras et~al.(2015)Gras, Leclerc, Hild, Roux and
  Schneider}]{gras_identification_2015}
\bibinfo{author}{Gras, R.}, \bibinfo{author}{Leclerc, H.},
  \bibinfo{author}{Hild, F.}, \bibinfo{author}{Roux, S.},
  \bibinfo{author}{Schneider, J.}, \bibinfo{year}{2015}.
\newblock \bibinfo{title}{Identification of a set of macroscopic elastic
  parameters in a {3D} woven composite: {Uncertainty} analysis and
  regularization}.
\newblock \bibinfo{journal}{International Journal of Solids and Structures}
  \bibinfo{volume}{55}, \bibinfo{pages}{2--16}.
\newblock \URLprefix
  \url{https://www.sciencedirect.com/science/article/pii/S0020768313004964},
  \DOIprefix\doi{10.1016/j.ijsolstr.2013.12.023}.
\bibitem[{Greaves et~al.(2011)Greaves, Greer, Lakes and
  Rouxel}]{greaves_poissons_2011}
\bibinfo{author}{Greaves, G.N.}, \bibinfo{author}{Greer, A.L.},
  \bibinfo{author}{Lakes, R.S.}, \bibinfo{author}{Rouxel, T.},
  \bibinfo{year}{2011}.
\newblock \bibinfo{title}{Poisson's ratio and modern materials}.
\newblock \bibinfo{journal}{Nature Materials} \bibinfo{volume}{10},
  \bibinfo{pages}{823--837}.
\newblock \URLprefix \url{https://www.nature.com/articles/nmat3134},
  \DOIprefix\doi{10.1038/nmat3134}. \bibinfo{note}{number: 11 Publisher: Nature
  Publishing Group}.
\bibitem[{Grédiac(1989)}]{grediac_principle_1989}
\bibinfo{author}{Grédiac, M.}, \bibinfo{year}{1989}.
\newblock \bibinfo{title}{Principle of virtual work and identification}.
\newblock \bibinfo{journal}{Comptes Rendus de L Academie des Sciences Serie Ii}
  , \bibinfo{pages}{1--5}.
\bibitem[{Grédiac and Pierron(2006)}]{grediac_applying_2006}
\bibinfo{author}{Grédiac, M.}, \bibinfo{author}{Pierron, F.},
  \bibinfo{year}{2006}.
\newblock \bibinfo{title}{Applying the {Virtual} {Fields} {Method} to the
  identification of elasto-plastic constitutive parameters}.
\newblock \bibinfo{journal}{International Journal of Plasticity}
  \bibinfo{volume}{22}, \bibinfo{pages}{602--627}.
\newblock \URLprefix
  \url{https://linkinghub.elsevier.com/retrieve/pii/S0749641905000896},
  \DOIprefix\doi{10.1016/j.ijplas.2005.04.007}.
\bibitem[{Grédiac et~al.(2008)Grédiac, Pierron, Avril and
  Toussaint}]{grediac_virtual_2008}
\bibinfo{author}{Grédiac, M.}, \bibinfo{author}{Pierron, F.},
  \bibinfo{author}{Avril, S.}, \bibinfo{author}{Toussaint, E.},
  \bibinfo{year}{2008}.
\newblock \bibinfo{title}{The {Virtual} {Fields} {Method} for {Extracting}
  {Constitutive} {Parameters} {From} {Full}-{Field} {Measurements}: a
  {Review}}.
\newblock \bibinfo{journal}{Strain} \bibinfo{volume}{42},
  \bibinfo{pages}{233--253}.
\newblock \URLprefix
  \url{http://doi.wiley.com/10.1111/j.1475-1305.2006.tb01504.x},
  \DOIprefix\doi{10.1111/j.1475-1305.2006.tb01504.x}.
\bibitem[{Guseinov et~al.(2020)Guseinov, McMahan, Pérez, Daraio and
  Bickel}]{guseinov_programming_2020}
\bibinfo{author}{Guseinov, R.}, \bibinfo{author}{McMahan, C.},
  \bibinfo{author}{Pérez, J.}, \bibinfo{author}{Daraio, C.},
  \bibinfo{author}{Bickel, B.}, \bibinfo{year}{2020}.
\newblock \bibinfo{title}{Programming temporal morphing of self-actuated
  shells}.
\newblock \bibinfo{journal}{Nature Communications} \bibinfo{volume}{11},
  \bibinfo{pages}{237}.
\newblock \URLprefix \url{https://www.nature.com/articles/s41467-019-14015-2},
  \DOIprefix\doi{10.1038/s41467-019-14015-2}. \bibinfo{note}{number: 1
  Publisher: Nature Publishing Group}.
\bibitem[{Hild and Roux(2012)}]{hild_comparison_2012}
\bibinfo{author}{Hild, F.}, \bibinfo{author}{Roux, S.}, \bibinfo{year}{2012}.
\newblock \bibinfo{title}{Comparison of {Local} and {Global} {Approaches} to
  {Digital} {Image} {Correlation}}.
\newblock \bibinfo{journal}{Experimental Mechanics} \bibinfo{volume}{52},
  \bibinfo{pages}{1503--1519}.
\newblock \URLprefix \url{http://link.springer.com/10.1007/s11340-012-9603-7},
  \DOIprefix\doi{10.1007/s11340-012-9603-7}.
\bibitem[{Huang et~al.(2020)Huang, Xu, Farhat and Darve}]{huang_learning_2020}
\bibinfo{author}{Huang, D.Z.}, \bibinfo{author}{Xu, K.},
  \bibinfo{author}{Farhat, C.}, \bibinfo{author}{Darve, E.},
  \bibinfo{year}{2020}.
\newblock \bibinfo{title}{Learning constitutive relations from indirect
  observations using deep neural networks}.
\newblock \bibinfo{journal}{Journal of Computational Physics}
  \bibinfo{volume}{416}, \bibinfo{pages}{109491}.
\newblock \URLprefix
  \url{https://linkinghub.elsevier.com/retrieve/pii/S0021999120302655},
  \DOIprefix\doi{10.1016/j.jcp.2020.109491}.
\bibitem[{Joshi et~al.(2022)Joshi, Thakolkaran, Zheng, Escande, Flaschel,
  De~Lorenzis and Kumar}]{joshi_bayesian-euclid_2022}
\bibinfo{author}{Joshi, A.}, \bibinfo{author}{Thakolkaran, P.},
  \bibinfo{author}{Zheng, Y.}, \bibinfo{author}{Escande, M.},
  \bibinfo{author}{Flaschel, M.}, \bibinfo{author}{De~Lorenzis, L.},
  \bibinfo{author}{Kumar, S.}, \bibinfo{year}{2022}.
\newblock \bibinfo{title}{Bayesian-{EUCLID}: {Discovering} hyperelastic
  material laws with uncertainties}.
\newblock \bibinfo{journal}{Computer Methods in Applied Mechanics and
  Engineering} \bibinfo{volume}{398}, \bibinfo{pages}{115225}.
\newblock \URLprefix
  \url{https://linkinghub.elsevier.com/retrieve/pii/S0045782522003681},
  \DOIprefix\doi{10.1016/j.cma.2022.115225}.
\bibitem[{Kadic et~al.(2012)Kadic, Bückmann, Stenger, Thiel and
  Wegener}]{kadic_practicability_2012}
\bibinfo{author}{Kadic, M.}, \bibinfo{author}{Bückmann, T.},
  \bibinfo{author}{Stenger, N.}, \bibinfo{author}{Thiel, M.},
  \bibinfo{author}{Wegener, M.}, \bibinfo{year}{2012}.
\newblock \bibinfo{title}{On the practicability of pentamode mechanical
  metamaterials}.
\newblock \bibinfo{journal}{Applied Physics Letters} \bibinfo{volume}{100},
  \bibinfo{pages}{191901}.
\newblock \URLprefix \url{https://aip.scitation.org/doi/10.1063/1.4709436},
  \DOIprefix\doi{10.1063/1.4709436}. \bibinfo{note}{publisher: American
  Institute of Physics}.
\bibitem[{Kaina et~al.(2015)Kaina, Lemoult, Fink and
  Lerosey}]{kaina_negative_2015}
\bibinfo{author}{Kaina, N.}, \bibinfo{author}{Lemoult, F.},
  \bibinfo{author}{Fink, M.}, \bibinfo{author}{Lerosey, G.},
  \bibinfo{year}{2015}.
\newblock \bibinfo{title}{Negative refractive index and acoustic superlens from
  multiple scattering in single negative metamaterials}.
\newblock \bibinfo{journal}{Nature} \bibinfo{volume}{525},
  \bibinfo{pages}{77--81}.
\newblock \URLprefix \url{https://www.nature.com/articles/nature14678},
  \DOIprefix\doi{10.1038/nature14678}. \bibinfo{note}{number: 7567 Publisher:
  Nature Publishing Group}.
\bibitem[{Karathanasopoulos et~al.(2020)Karathanasopoulos, Dos~Reis,
  Diamantopoulou and Ganghoffer}]{karathanasopoulos_mechanics_2020}
\bibinfo{author}{Karathanasopoulos, N.}, \bibinfo{author}{Dos~Reis, F.},
  \bibinfo{author}{Diamantopoulou, M.}, \bibinfo{author}{Ganghoffer, J.F.},
  \bibinfo{year}{2020}.
\newblock \bibinfo{title}{Mechanics of beams made from chiral metamaterials:
  {Tuning} deflections through normal-shear strain couplings}.
\newblock \bibinfo{journal}{Materials \& Design} \bibinfo{volume}{189},
  \bibinfo{pages}{108520}.
\newblock \URLprefix
  \url{https://www.sciencedirect.com/science/article/pii/S0264127520300538},
  \DOIprefix\doi{10.1016/j.matdes.2020.108520}.
\bibitem[{Kim et~al.(2020)Kim, Kim and Lee}]{kim_virtual_2020}
\bibinfo{author}{Kim, C.}, \bibinfo{author}{Kim, J.H.}, \bibinfo{author}{Lee,
  M.G.}, \bibinfo{year}{2020}.
\newblock \bibinfo{title}{A virtual fields method for identifying anisotropic
  elastic constants of fiber reinforced composites using a single tension test:
  {Theory} and validation}.
\newblock \bibinfo{journal}{Composites Part B: Engineering}
  \bibinfo{volume}{200}, \bibinfo{pages}{108338}.
\newblock \URLprefix
  \url{https://www.sciencedirect.com/science/article/pii/S1359836820333874},
  \DOIprefix\doi{10.1016/j.compositesb.2020.108338}.
\bibitem[{Kulagin et~al.(2020)Kulagin, Beygelzimer, Estrin, Schumilin and
  Gumbsch}]{kulagin_architectured_2020}
\bibinfo{author}{Kulagin, R.}, \bibinfo{author}{Beygelzimer, Y.},
  \bibinfo{author}{Estrin, Y.}, \bibinfo{author}{Schumilin, A.},
  \bibinfo{author}{Gumbsch, P.}, \bibinfo{year}{2020}.
\newblock \bibinfo{title}{Architectured {Lattice} {Materials} with {Tunable}
  {Anisotropy}: {Design} and {Analysis} of the {Material} {Property} {Space}
  with the {Aid} of {Machine} {Learning}}.
\newblock \bibinfo{journal}{Advanced Engineering Materials}
  \bibinfo{volume}{22}, \bibinfo{pages}{2001069}.
\newblock \URLprefix
  \url{https://onlinelibrary.wiley.com/doi/10.1002/adem.202001069},
  \DOIprefix\doi{10.1002/adem.202001069}.
\bibitem[{Kumar et~al.(2020)Kumar, Tan, Zheng and Kochmann}]{Kumar2020}
\bibinfo{author}{Kumar, S.}, \bibinfo{author}{Tan, S.}, \bibinfo{author}{Zheng,
  L.}, \bibinfo{author}{Kochmann, D.M.}, \bibinfo{year}{2020}.
\newblock \bibinfo{title}{Inverse-designed spinodoid metamaterials}.
\newblock \bibinfo{journal}{npj Computational Materials} \bibinfo{volume}{6}.
\newblock \URLprefix \url{https://doi.org/10.1038/s41524-020-0341-6},
  \DOIprefix\doi{10.1038/s41524-020-0341-6}.
\bibitem[{Lee et~al.(2016)Lee, Lee, Ma and Kim}]{lee_effective_2016}
\bibinfo{author}{Lee, H.J.}, \bibinfo{author}{Lee, H.S.}, \bibinfo{author}{Ma,
  P.S.}, \bibinfo{author}{Kim, Y.Y.}, \bibinfo{year}{2016}.
\newblock \bibinfo{title}{Effective material parameter retrieval of anisotropic
  elastic metamaterials with inherent nonlocality}.
\newblock \bibinfo{journal}{Journal of Applied Physics} \bibinfo{volume}{120},
  \bibinfo{pages}{104902}.
\newblock \URLprefix \url{https://aip.scitation.org/doi/10.1063/1.4962274},
  \DOIprefix\doi{10.1063/1.4962274}. \bibinfo{note}{publisher: American
  Institute of Physics}.
\bibitem[{Lee et~al.(2012)Lee, Singer and
  Thomas}]{lee_micro-nanostructured_2012}
\bibinfo{author}{Lee, J.H.}, \bibinfo{author}{Singer, J.P.},
  \bibinfo{author}{Thomas, E.L.}, \bibinfo{year}{2012}.
\newblock \bibinfo{title}{Micro-/{Nanostructured} {Mechanical}
  {Metamaterials}}.
\newblock \bibinfo{journal}{Advanced Materials} \bibinfo{volume}{24},
  \bibinfo{pages}{4782--4810}.
\newblock \URLprefix
  \url{https://onlinelibrary.wiley.com/doi/abs/10.1002/adma.201201644},
  \DOIprefix\doi{10.1002/adma.201201644}. \bibinfo{note}{\_eprint:
  https://onlinelibrary.wiley.com/doi/pdf/10.1002/adma.201201644}.
\bibitem[{Liu et~al.(2020)Liu, Tao, Du, Yu and Xu}]{liu_learning_2020}
\bibinfo{author}{Liu, X.}, \bibinfo{author}{Tao, F.}, \bibinfo{author}{Du, H.},
  \bibinfo{author}{Yu, W.}, \bibinfo{author}{Xu, K.}, \bibinfo{year}{2020}.
\newblock \bibinfo{title}{Learning {Nonlinear} {Constitutive} {Laws} {Using}
  {Neural} {Network} {Models} {Based} on {Indirectly} {Measurable} {Data}}.
\newblock \bibinfo{journal}{Journal of Applied Mechanics} \bibinfo{volume}{87},
  \bibinfo{pages}{081003}.
\newblock \URLprefix
  \url{https://asmedigitalcollection.asme.org/appliedmechanics/article/doi/10.1115/1.4047036/1083320/Learning-Nonlinear-Constitutive-Laws-Using-Neural},
  \DOIprefix\doi{10.1115/1.4047036}.
\bibitem[{Man and Furukawa(2011)}]{man_neural_2011}
\bibinfo{author}{Man, H.}, \bibinfo{author}{Furukawa, T.},
  \bibinfo{year}{2011}.
\newblock \bibinfo{title}{Neural network constitutive modelling for non-linear
  characterization of anisotropic materials}.
\newblock \bibinfo{journal}{International Journal for Numerical Methods in
  Engineering} \bibinfo{volume}{85}, \bibinfo{pages}{939--957}.
\newblock \URLprefix
  \url{https://onlinelibrary.wiley.com/doi/10.1002/nme.2999},
  \DOIprefix\doi{10.1002/nme.2999}.
\bibitem[{Mao et~al.(2020)Mao, Rumpler, Gaborit, Göransson, Kennedy, O'Connor,
  Trimble and Rice}]{mao_twist_2020}
\bibinfo{author}{Mao, H.}, \bibinfo{author}{Rumpler, R.},
  \bibinfo{author}{Gaborit, M.}, \bibinfo{author}{Göransson, P.},
  \bibinfo{author}{Kennedy, J.}, \bibinfo{author}{O'Connor, D.},
  \bibinfo{author}{Trimble, D.}, \bibinfo{author}{Rice, H.},
  \bibinfo{year}{2020}.
\newblock \bibinfo{title}{Twist, tilt and stretch: {From} isometric {Kelvin}
  cells to anisotropic cellular materials}.
\newblock \bibinfo{journal}{Materials \& Design} \bibinfo{volume}{193},
  \bibinfo{pages}{108855}.
\newblock \URLprefix
  \url{https://linkinghub.elsevier.com/retrieve/pii/S0264127520303890},
  \DOIprefix\doi{10.1016/j.matdes.2020.108855}.
\bibitem[{Marek et~al.(2017)Marek, Davis and
  Pierron}]{marek_sensitivity-based_2017}
\bibinfo{author}{Marek, A.}, \bibinfo{author}{Davis, F.M.},
  \bibinfo{author}{Pierron, F.}, \bibinfo{year}{2017}.
\newblock \bibinfo{title}{Sensitivity-based virtual fields for the non-linear
  virtual fields method}.
\newblock \bibinfo{journal}{Computational Mechanics} \bibinfo{volume}{60},
  \bibinfo{pages}{409--431}.
\newblock \URLprefix \url{http://link.springer.com/10.1007/s00466-017-1411-6},
  \DOIprefix\doi{10.1007/s00466-017-1411-6}.
\bibitem[{Marino et~al.(2023)Marino, Flaschel, Kumar and
  De~Lorenzis}]{marino_automated_2023}
\bibinfo{author}{Marino, E.}, \bibinfo{author}{Flaschel, M.},
  \bibinfo{author}{Kumar, S.}, \bibinfo{author}{De~Lorenzis, L.},
  \bibinfo{year}{2023}.
\newblock \bibinfo{title}{Automated identification of linear viscoelastic
  constitutive laws with {EUCLID}}.
\newblock \bibinfo{journal}{Mechanics of Materials} \bibinfo{volume}{181},
  \bibinfo{pages}{104643}.
\newblock \URLprefix
  \url{https://linkinghub.elsevier.com/retrieve/pii/S0167663623000893},
  \DOIprefix\doi{10.1016/j.mechmat.2023.104643}.
\bibitem[{Milton and Cherkaev(1995)}]{milton_which_1995}
\bibinfo{author}{Milton, G.W.}, \bibinfo{author}{Cherkaev, A.V.},
  \bibinfo{year}{1995}.
\newblock \bibinfo{title}{Which {Elasticity} {Tensors} are {Realizable}?}
\newblock \bibinfo{journal}{Journal of Engineering Materials and Technology}
  \bibinfo{volume}{117}, \bibinfo{pages}{483--493}.
\newblock \URLprefix \url{https://doi.org/10.1115/1.2804743},
  \DOIprefix\doi{10.1115/1.2804743}.
\bibitem[{Mindlin and Eshel(1968)}]{mindlin_first_1968}
\bibinfo{author}{Mindlin, R.D.}, \bibinfo{author}{Eshel, N.N.},
  \bibinfo{year}{1968}.
\newblock \bibinfo{title}{On first strain-gradient theories in linear
  elasticity}.
\newblock \bibinfo{journal}{International Journal of Solids and Structures}
  \bibinfo{volume}{4}, \bibinfo{pages}{109--124}.
\newblock \URLprefix
  \url{https://www.sciencedirect.com/science/article/pii/002076836890036X},
  \DOIprefix\doi{10.1016/0020-7683(68)90036-X}.
\bibitem[{Ni et~al.(2019)Ni, Guo, Li, Huang, Zhang and Rogers}]{ni_2d_2019}
\bibinfo{author}{Ni, X.}, \bibinfo{author}{Guo, X.}, \bibinfo{author}{Li, J.},
  \bibinfo{author}{Huang, Y.}, \bibinfo{author}{Zhang, Y.},
  \bibinfo{author}{Rogers, J.A.}, \bibinfo{year}{2019}.
\newblock \bibinfo{title}{{2D} {Mechanical} {Metamaterials} with {Widely}
  {Tunable} {Unusual} {Modes} of {Thermal} {Expansion}}.
\newblock \bibinfo{journal}{Advanced Materials} \bibinfo{volume}{31},
  \bibinfo{pages}{1905405}.
\newblock \URLprefix
  \url{https://onlinelibrary.wiley.com/doi/abs/10.1002/adma.201905405},
  \DOIprefix\doi{10.1002/adma.201905405}. \bibinfo{note}{\_eprint:
  https://onlinelibrary.wiley.com/doi/pdf/10.1002/adma.201905405}.
\bibitem[{Pierron(2023)}]{pierron_material_nodate}
\bibinfo{author}{Pierron, F.}, \bibinfo{year}{2023}.
\newblock \bibinfo{title}{Material testing 2.0: A brief review}.
\newblock \bibinfo{journal}{Strain} \bibinfo{volume}{n/a},
  \bibinfo{pages}{e12434}.
\newblock \URLprefix
  \url{https://onlinelibrary.wiley.com/doi/abs/10.1111/str.12434},
  \DOIprefix\doi{10.1111/str.12434}.
\bibitem[{Pierron et~al.(2010)Pierron, Avril and Tran}]{pierron_extension_2010}
\bibinfo{author}{Pierron, F.}, \bibinfo{author}{Avril, S.},
  \bibinfo{author}{Tran, V.T.}, \bibinfo{year}{2010}.
\newblock \bibinfo{title}{Extension of the virtual fields method to
  elasto-plastic material identification with cyclic loads and kinematic
  hardening}.
\newblock \bibinfo{journal}{International Journal of Solids and Structures}
  \bibinfo{volume}{47}, \bibinfo{pages}{2993--3010}.
\newblock \URLprefix
  \url{https://www.sciencedirect.com/science/article/pii/S002076831000243X},
  \DOIprefix\doi{10.1016/j.ijsolstr.2010.06.022}.
\bibitem[{Pierron and Grédiac(2012)}]{pierron_virtual_2012}
\bibinfo{author}{Pierron, F.}, \bibinfo{author}{Grédiac, M.},
  \bibinfo{year}{2012}.
\newblock \bibinfo{title}{The {Virtual} {Fields} {Method}}.
\newblock \bibinfo{publisher}{Springer New York}, \bibinfo{address}{New York,
  NY}.
\newblock \URLprefix \url{http://link.springer.com/10.1007/978-1-4614-1824-5},
  \DOIprefix\doi{10.1007/978-1-4614-1824-5}.
\bibitem[{Podestá et~al.(2019)Podestá, Méndez, Toro and
  Huespe}]{podesta_symmetry_2019}
\bibinfo{author}{Podestá, J.}, \bibinfo{author}{Méndez, C.},
  \bibinfo{author}{Toro, S.}, \bibinfo{author}{Huespe, A.},
  \bibinfo{year}{2019}.
\newblock \bibinfo{title}{Symmetry considerations for topology design in the
  elastic inverse homogenization problem}.
\newblock \bibinfo{journal}{Journal of the Mechanics and Physics of Solids}
  \bibinfo{volume}{128}, \bibinfo{pages}{54--78}.
\newblock \URLprefix
  \url{https://linkinghub.elsevier.com/retrieve/pii/S0022509618305210},
  \DOIprefix\doi{10.1016/j.jmps.2019.03.018}.
\bibitem[{Promma et~al.(2009)Promma, Raka, Grédiac, Toussaint, Le~Cam,
  Balandraud and Hild}]{promma_application_2009}
\bibinfo{author}{Promma, N.}, \bibinfo{author}{Raka, B.},
  \bibinfo{author}{Grédiac, M.}, \bibinfo{author}{Toussaint, E.},
  \bibinfo{author}{Le~Cam, J.B.}, \bibinfo{author}{Balandraud, X.},
  \bibinfo{author}{Hild, F.}, \bibinfo{year}{2009}.
\newblock \bibinfo{title}{Application of the virtual fields method to
  mechanical characterization of elastomeric materials}.
\newblock \bibinfo{journal}{International Journal of Solids and Structures}
  \bibinfo{volume}{46}, \bibinfo{pages}{698--715}.
\newblock \URLprefix
  \url{https://www.sciencedirect.com/science/article/pii/S0020768308003946},
  \DOIprefix\doi{10.1016/j.ijsolstr.2008.09.025}.
\bibitem[{Risso et~al.(2021)Risso, Sakovsky and
  Ermanni}]{risso_instability-driven_2021}
\bibinfo{author}{Risso, G.}, \bibinfo{author}{Sakovsky, M.},
  \bibinfo{author}{Ermanni, P.}, \bibinfo{year}{2021}.
\newblock \bibinfo{title}{Instability-driven shape forming of fiber reinforced
  polymer frames}.
\newblock \bibinfo{journal}{Composite Structures} \bibinfo{volume}{268},
  \bibinfo{pages}{113946}.
\newblock \URLprefix
  \url{https://www.sciencedirect.com/science/article/pii/S0263822321004062},
  \DOIprefix\doi{10.1016/j.compstruct.2021.113946}.
\bibitem[{Rosakis and Ravi-Chandar(1986)}]{rosakis_crack-tip_1986}
\bibinfo{author}{Rosakis, A.J.}, \bibinfo{author}{Ravi-Chandar, K.},
  \bibinfo{year}{1986}.
\newblock \bibinfo{title}{On crack-tip stress state: {An} experimental
  evaluation of three-dimensional effects}.
\newblock \bibinfo{journal}{International Journal of Solids and Structures}
  \bibinfo{volume}{22}, \bibinfo{pages}{121--134}.
\newblock \URLprefix
  \url{https://www.sciencedirect.com/science/article/pii/0020768386900028},
  \DOIprefix\doi{10.1016/0020-7683(86)90002-8}.
\bibitem[{Roux and Hild(2020)}]{roux_optimal_2020}
\bibinfo{author}{Roux, S.}, \bibinfo{author}{Hild, F.}, \bibinfo{year}{2020}.
\newblock \bibinfo{title}{Optimal procedure for the identification of
  constitutive parameters from experimentally measured displacement fields}.
\newblock \bibinfo{journal}{International Journal of Solids and Structures}
  \bibinfo{volume}{184}, \bibinfo{pages}{14--23}.
\newblock \URLprefix
  \url{https://linkinghub.elsevier.com/retrieve/pii/S0020768318304542},
  \DOIprefix\doi{10.1016/j.ijsolstr.2018.11.008}.
\bibitem[{Rychlewski(1984)}]{rychlewski_hookes_1984}
\bibinfo{author}{Rychlewski, J.}, \bibinfo{year}{1984}.
\newblock \bibinfo{title}{On {Hooke}'s law}.
\newblock \bibinfo{journal}{Journal of Applied Mathematics and Mechanics}
  \bibinfo{volume}{48}, \bibinfo{pages}{303--314}.
\newblock \URLprefix
  \url{https://www.sciencedirect.com/science/article/pii/0021892884901370},
  \DOIprefix\doi{10.1016/0021-8928(84)90137-0}.
\bibitem[{Schittny et~al.(2013)Schittny, Bückmann, Kadic and
  Wegener}]{schittny_elastic_2013}
\bibinfo{author}{Schittny, R.}, \bibinfo{author}{Bückmann, T.},
  \bibinfo{author}{Kadic, M.}, \bibinfo{author}{Wegener, M.},
  \bibinfo{year}{2013}.
\newblock \bibinfo{title}{Elastic measurements on macroscopic three-dimensional
  pentamode metamaterials}.
\newblock \bibinfo{journal}{Applied Physics Letters} \bibinfo{volume}{103},
  \bibinfo{pages}{231905}.
\newblock \URLprefix \url{https://aip.scitation.org/doi/10.1063/1.4838663},
  \DOIprefix\doi{10.1063/1.4838663}. \bibinfo{note}{publisher: American
  Institute of Physics}.
\bibitem[{Soyarslan et~al.(2018)Soyarslan, Bargmann, Pradas and
  Weissm\"{u}ller}]{Soyarslan2018}
\bibinfo{author}{Soyarslan, C.}, \bibinfo{author}{Bargmann, S.},
  \bibinfo{author}{Pradas, M.}, \bibinfo{author}{Weissm\"{u}ller, J.},
  \bibinfo{year}{2018}.
\newblock \bibinfo{title}{3d stochastic bicontinuous microstructures:
  Generation, topology and elasticity}.
\newblock \bibinfo{journal}{Acta Materialia} \bibinfo{volume}{149},
  \bibinfo{pages}{326--340}.
\newblock \URLprefix \url{https://doi.org/10.1016/j.actamat.2018.01.005},
  \DOIprefix\doi{10.1016/j.actamat.2018.01.005}.
\bibitem[{Surjadi et~al.(2019)Surjadi, Gao, Du, Li, Xiong, Fang and
  Lu}]{surjadi_mechanical_2019}
\bibinfo{author}{Surjadi, J.U.}, \bibinfo{author}{Gao, L.},
  \bibinfo{author}{Du, H.}, \bibinfo{author}{Li, X.}, \bibinfo{author}{Xiong,
  X.}, \bibinfo{author}{Fang, N.X.}, \bibinfo{author}{Lu, Y.},
  \bibinfo{year}{2019}.
\newblock \bibinfo{title}{Mechanical {Metamaterials} and {Their} {Engineering}
  {Applications}}.
\newblock \bibinfo{journal}{Advanced Engineering Materials}
  \bibinfo{volume}{21}, \bibinfo{pages}{1800864}.
\newblock \URLprefix
  \url{https://onlinelibrary.wiley.com/doi/abs/10.1002/adem.201800864},
  \DOIprefix\doi{10.1002/adem.201800864}. \bibinfo{note}{\_eprint:
  https://onlinelibrary.wiley.com/doi/pdf/10.1002/adem.201800864}.
\bibitem[{Sutton et~al.(2009)Sutton, Orteu and Schreier}]{sutton_image_2009}
\bibinfo{author}{Sutton, M.A.}, \bibinfo{author}{Orteu, J.J.},
  \bibinfo{author}{Schreier, H.}, \bibinfo{year}{2009}.
\newblock \bibinfo{title}{Image correlation for shape, motion and deformation
  measurements: basic concepts, theory and applications}.
\newblock \bibinfo{publisher}{Springer Science \& Business Media}.
\bibitem[{Thakolkaran et~al.(2022)Thakolkaran, Joshi, Zheng, Flaschel,
  De~Lorenzis and Kumar}]{thakolkaran_nn-euclid_2022}
\bibinfo{author}{Thakolkaran, P.}, \bibinfo{author}{Joshi, A.},
  \bibinfo{author}{Zheng, Y.}, \bibinfo{author}{Flaschel, M.},
  \bibinfo{author}{De~Lorenzis, L.}, \bibinfo{author}{Kumar, S.},
  \bibinfo{year}{2022}.
\newblock \bibinfo{title}{{NN}-{EUCLID}: {Deep}-learning hyperelasticity
  without stress data}.
\newblock \bibinfo{journal}{Journal of the Mechanics and Physics of Solids}
  \bibinfo{volume}{169}, \bibinfo{pages}{105076}.
\newblock \URLprefix
  \url{https://linkinghub.elsevier.com/retrieve/pii/S0022509622002538},
  \DOIprefix\doi{10.1016/j.jmps.2022.105076}.
\bibitem[{Tibshirani(1996)}]{tibshirani_regression_1996}
\bibinfo{author}{Tibshirani, R.}, \bibinfo{year}{1996}.
\newblock \bibinfo{title}{Regression {Shrinkage} and {Selection} via the
  {Lasso}}.
\newblock \bibinfo{journal}{Journal of the Royal Statistical Society: Series B
  (Methodological)} \bibinfo{volume}{58}, \bibinfo{pages}{267--288}.
\newblock \URLprefix
  \url{http://doi.wiley.com/10.1111/j.2517-6161.1996.tb02080.x},
  \DOIprefix\doi{10.1111/j.2517-6161.1996.tb02080.x}.
\bibitem[{Ting and Chi-Tsai(1996)}]{ting1996anisotropic}
\bibinfo{author}{Ting}, \bibinfo{author}{Chi-Tsai, T.}, \bibinfo{year}{1996}.
\newblock \bibinfo{title}{\href{
  https://doi.org/10.1002/zamm.19970770617}{Anisotropic elasticity: theory and
  applications}}.
\newblock \bibinfo{number}{45}, \bibinfo{publisher}{Oxford University Press on
  Demand}.
\bibitem[{Wu et~al.(2019)Wu, Hu, Qian, Liao, Xu and Berto}]{wu_mechanical_2019}
\bibinfo{author}{Wu, W.}, \bibinfo{author}{Hu, W.}, \bibinfo{author}{Qian, G.},
  \bibinfo{author}{Liao, H.}, \bibinfo{author}{Xu, X.}, \bibinfo{author}{Berto,
  F.}, \bibinfo{year}{2019}.
\newblock \bibinfo{title}{Mechanical design and multifunctional applications of
  chiral mechanical metamaterials: {A} review}.
\newblock \bibinfo{journal}{Materials \& Design} \bibinfo{volume}{180},
  \bibinfo{pages}{107950}.
\newblock \URLprefix
  \url{https://www.sciencedirect.com/science/article/pii/S0264127519303880},
  \DOIprefix\doi{10.1016/j.matdes.2019.107950}.
\bibitem[{Yang et~al.(2021)Yang, Timofeev, Abali, Li and
  Müller}]{yang_verification_2021}
\bibinfo{author}{Yang, H.}, \bibinfo{author}{Timofeev, D.},
  \bibinfo{author}{Abali, B.E.}, \bibinfo{author}{Li, B.},
  \bibinfo{author}{Müller, W.H.}, \bibinfo{year}{2021}.
\newblock \bibinfo{title}{Verification of strain gradient elasticity
  computation by analytical solutions}.
\newblock \bibinfo{journal}{ZAMM - Journal of Applied Mathematics and Mechanics
  / Zeitschrift für Angewandte Mathematik und Mechanik} \URLprefix
  \url{https://onlinelibrary.wiley.com/doi/10.1002/zamm.202100023},
  \DOIprefix\doi{10.1002/zamm.202100023}.
\bibitem[{Yang et~al.(2019)Yang, Kweun and Kim}]{yang_monolayer_2019}
\bibinfo{author}{Yang, X.}, \bibinfo{author}{Kweun, M.}, \bibinfo{author}{Kim,
  Y.Y.}, \bibinfo{year}{2019}.
\newblock \bibinfo{title}{Monolayer metamaterial for full mode-converting
  transmission of elastic waves}.
\newblock \bibinfo{journal}{Applied Physics Letters} \bibinfo{volume}{115},
  \bibinfo{pages}{071901}.
\newblock \URLprefix \url{https://aip.scitation.org/doi/10.1063/1.5109758},
  \DOIprefix\doi{10.1063/1.5109758}. \bibinfo{note}{publisher: American
  Institute of Physics}.
\bibitem[{Yuan et~al.(2021)Yuan, Cui and Ju}]{yuan_micropolar_2021}
\bibinfo{author}{Yuan, Z.}, \bibinfo{author}{Cui, Z.}, \bibinfo{author}{Ju,
  J.}, \bibinfo{year}{2021}.
\newblock \bibinfo{title}{Micropolar homogenization of wavy tetra-chiral and
  tetra-achiral lattices to identify axial–shear coupling and directional
  negative {Poisson}'s ratio}.
\newblock \bibinfo{journal}{Materials \& Design} \bibinfo{volume}{201},
  \bibinfo{pages}{109483}.
\newblock \URLprefix
  \url{https://www.sciencedirect.com/science/article/pii/S0264127521000368},
  \DOIprefix\doi{10.1016/j.matdes.2021.109483}.
\bibitem[{Zadpoor(2016)}]{zadpoor_mechanical_2016}
\bibinfo{author}{Zadpoor, A.A.}, \bibinfo{year}{2016}.
\newblock \bibinfo{title}{Mechanical meta-materials}.
\newblock \bibinfo{journal}{Materials Horizons} \bibinfo{volume}{3},
  \bibinfo{pages}{371--381}.
\newblock \URLprefix \url{http://xlink.rsc.org/?DOI=C6MH00065G},
  \DOIprefix\doi{10.1039/C6MH00065G}.
\bibitem[{Zheng et~al.(2019)Zheng, Liu, Chen, Miao, Zhu and
  Hu}]{zheng_theory_2019}
\bibinfo{author}{Zheng, M.}, \bibinfo{author}{Liu, X.}, \bibinfo{author}{Chen,
  Y.}, \bibinfo{author}{Miao, H.}, \bibinfo{author}{Zhu, R.},
  \bibinfo{author}{Hu, G.}, \bibinfo{year}{2019}.
\newblock \bibinfo{title}{Theory and {Realization} of {Nonresonant}
  {Anisotropic} {Singly} {Polarized} {Solids} {Carrying} {Only} {Shear}
  {Waves}}.
\newblock \bibinfo{journal}{Physical Review Applied} \bibinfo{volume}{12},
  \bibinfo{pages}{014027}.
\newblock \URLprefix
  \url{https://link.aps.org/doi/10.1103/PhysRevApplied.12.014027},
  \DOIprefix\doi{10.1103/PhysRevApplied.12.014027}. \bibinfo{note}{publisher:
  American Physical Society}.
\bibitem[{Zheng et~al.(2020)Zheng, Park, Liu, Zhu, Hu and
  Kim}]{zheng_non-resonant_2020}
\bibinfo{author}{Zheng, M.}, \bibinfo{author}{Park, C.I.},
  \bibinfo{author}{Liu, X.}, \bibinfo{author}{Zhu, R.}, \bibinfo{author}{Hu,
  G.}, \bibinfo{author}{Kim, Y.Y.}, \bibinfo{year}{2020}.
\newblock \bibinfo{title}{Non-resonant metasurface for broadband elastic wave
  mode splitting}.
\newblock \bibinfo{journal}{Applied Physics Letters} \bibinfo{volume}{116},
  \bibinfo{pages}{171903}.
\newblock \URLprefix \url{https://aip.scitation.org/doi/10.1063/5.0005408},
  \DOIprefix\doi{10.1063/5.0005408}. \bibinfo{note}{publisher: American
  Institute of Physics}.

\end{thebibliography}
\clearpage
\begin{center}
\textbf{\large Supplementary Information}
\end{center}
\setcounter{equation}{0}
\setcounter{figure}{0}
\setcounter{table}{0}
\setcounter{page}{1}
\setcounter{section}{0}
\makeatletter
\renewcommand{\theequation}{S\arabic{equation}}
\renewcommand{\thefigure}{S\arabic{figure}}
\renewcommand{\bibnumfmt}[1]{[S#1]}
\renewcommand{\citenumfont}[1]{S#1}
\renewcommand{\thetable}{S\arabic{table}}
\renewcommand{\thesection}{S-\Roman{section}}
\section{Difference in displacement fields between homogeneous and heterogeneous materials}

\begin{figure}[!htb]
	\begin{center}
		\centering 
  \includegraphics[width=0.75\textwidth]{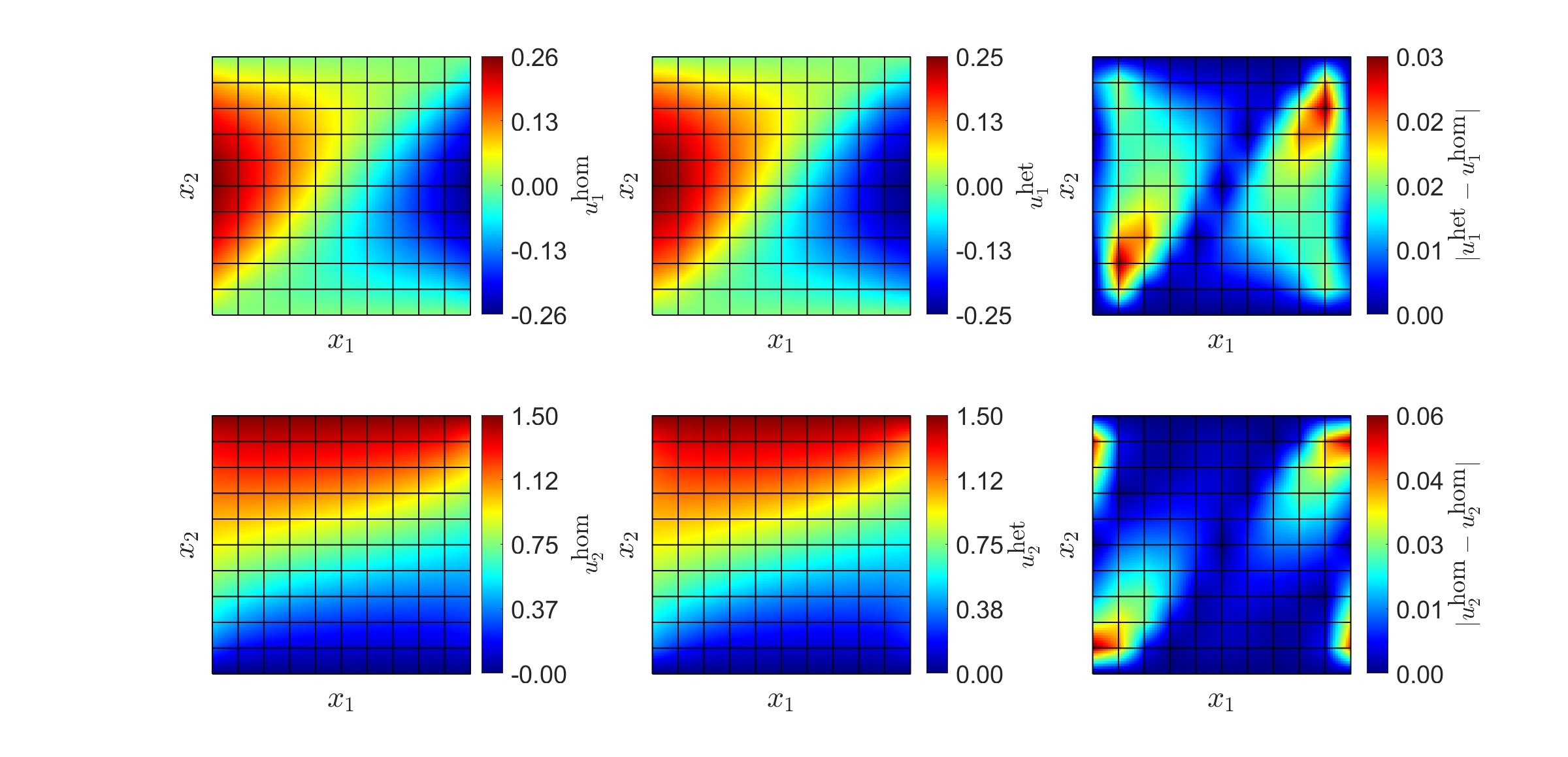}
		\caption{Comparison between the displacement fields obtained from finite element simulations of a homogeneous specimen (left) and a heterogeneous structure made of geometry \#2 (center). For the homogeneous specimen a finite element simulation using $10 \times 10$ bilinear quadrilateral elements was executed. The heterogeneous specimen was simulated using using $1000 \times 1000$ bilinear quadrilateral elements. Afterwards, the displacement data at the unit cell corners were extracted and interpolated with a bilinear polynomial for each unit cell, to allow for a comparison with the homogeneous specimen. The difference between the fields is shown on the right..}
		\label{fig:displacement_HOMvsHET_geom02}
	\end{center}
\end{figure}

\begin{figure}[!htb]
	\begin{center}
		\centering 
  \includegraphics[width=0.75\textwidth]{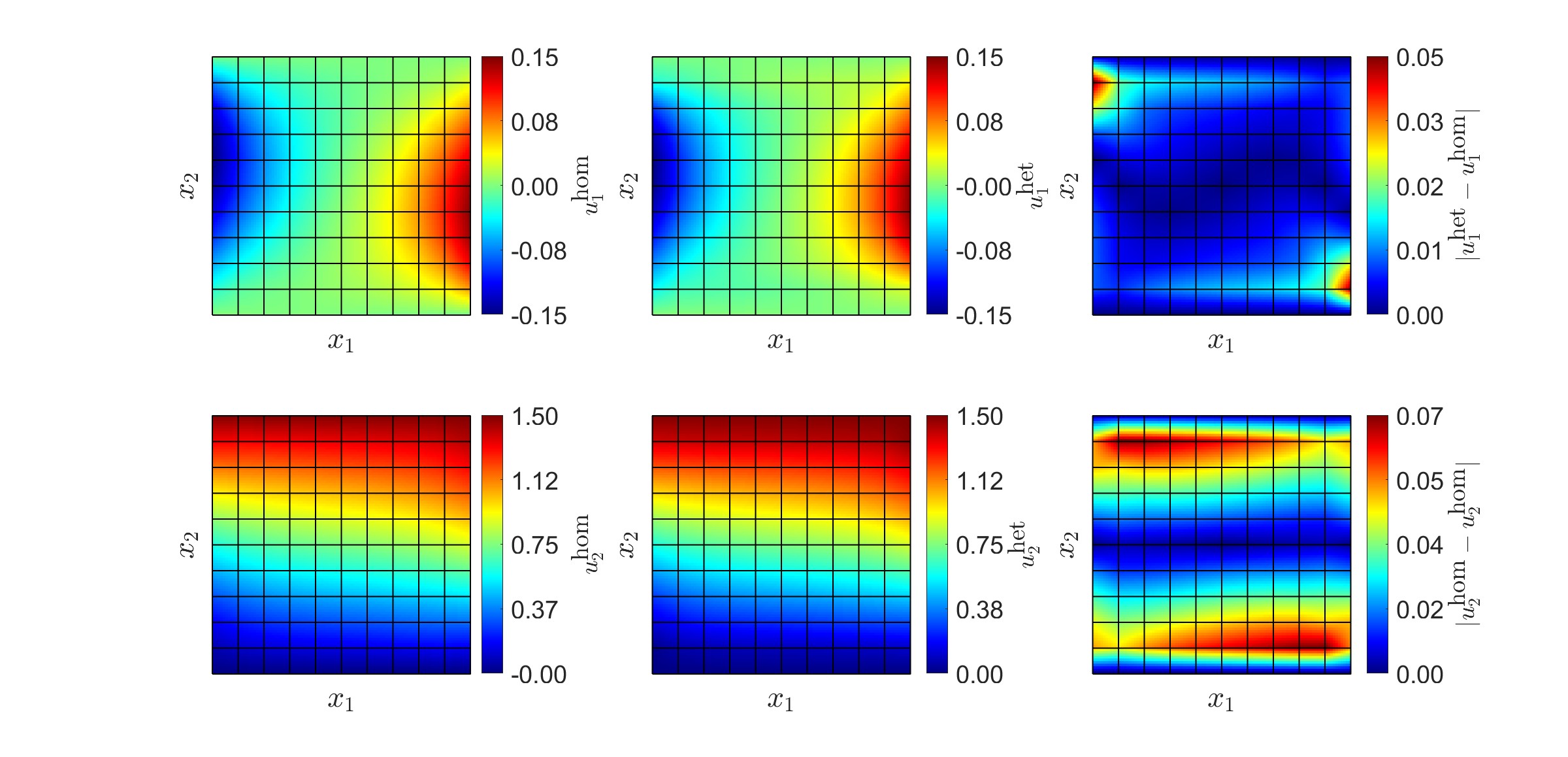}
		\caption{Comparison between the displacement fields obtained from finite element simulations of a homogeneous specimen (left) and a heterogeneous structure made of geometry \#3 (center). For the homogeneous specimen a finite element simulation using $10 \times 10$ bilinear quadrilateral elements was executed. The heterogeneous specimen was simulated using using $1000 \times 1000$ bilinear quadrilateral elements. Afterwards, the displacement data at the unit cell corners were extracted and interpolated with a bilinear polynomial for each unit cell, to allow for a comparison with the homogeneous specimen. The difference between the fields is shown on the right.}
		\label{fig:displacement_HOMvsHET_geom03}
	\end{center}
\end{figure}

\begin{figure}[!htb]
	\begin{center}
		\centering 
  	\includegraphics[width=0.75\textwidth]{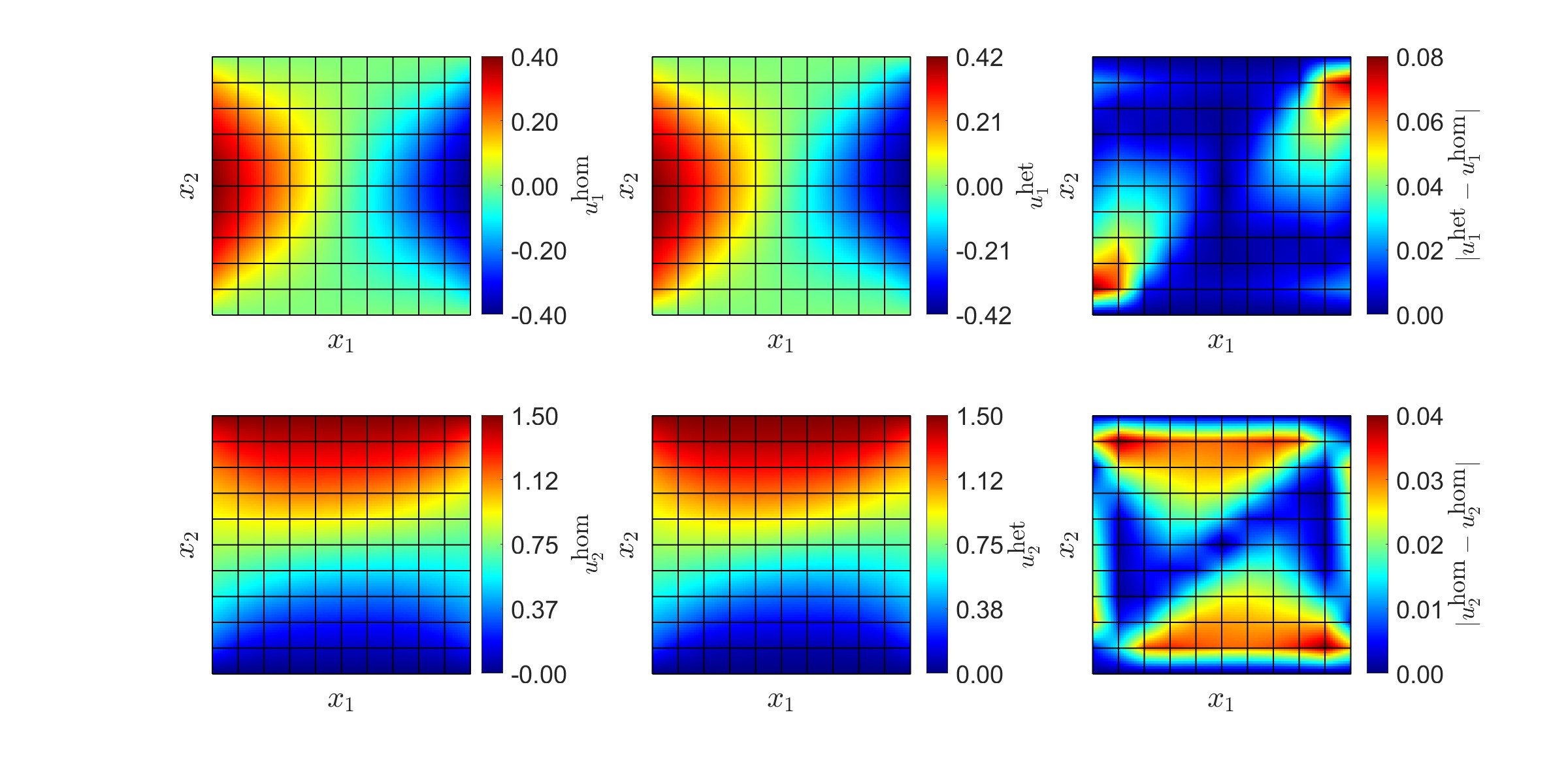}
		\caption{Comparison between the displacement fields obtained from finite element simulations of a homogeneous specimen (left) and a heterogeneous structure made of geometry \#4 (center). For the homogeneous specimen a finite element simulation using $10 \times 10$ bilinear quadrilateral elements was executed. The heterogeneous specimen was simulated using using $1000 \times 1000$ bilinear quadrilateral elements. Afterwards, the displacement data at the unit cell corners were extracted and interpolated with a bilinear polynomial for each unit cell, to allow for a comparison with the homogeneous specimen. The difference between the fields is shown on the right.}
		\label{fig:displacement_HOMvsHET_geom04}
	\end{center}
\end{figure}

\newpage
\section{Comparison of simulated and experimentally measured full-field displacement fields}
\begin{figure}[!htb]
	\begin{center}
		\centering 
		\includegraphics[width =0.99\textwidth]{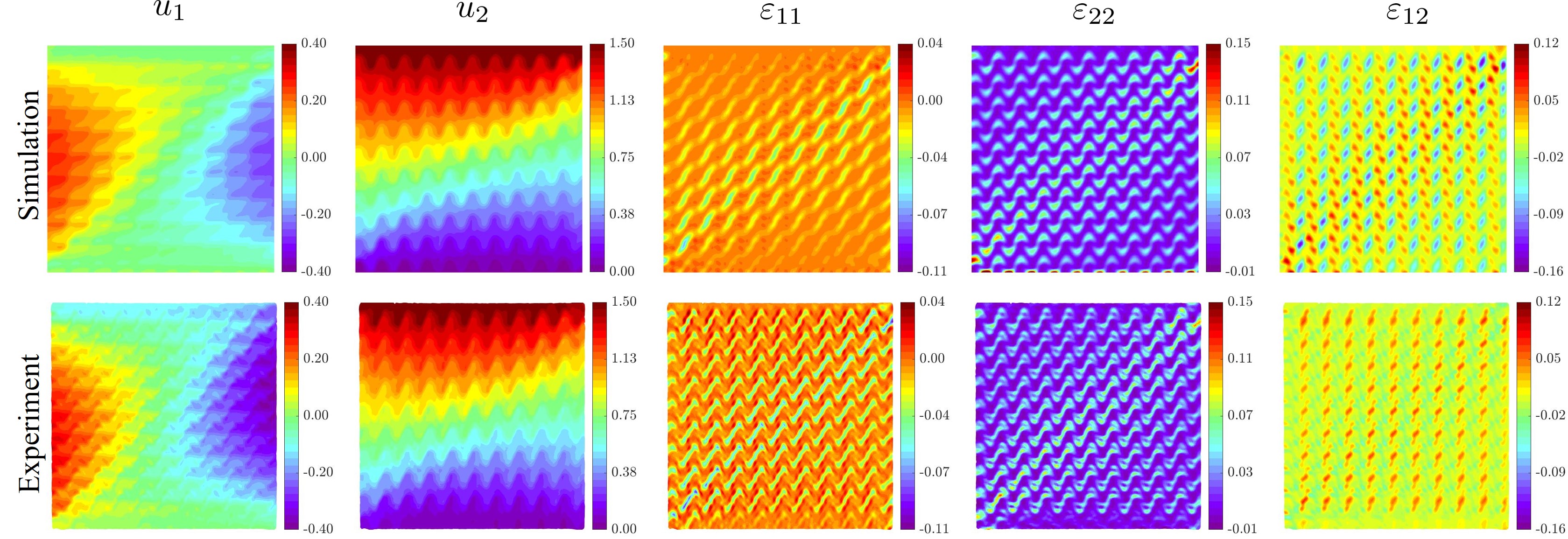}
		\caption{Comparison between numerical (top) and experimentally measured(bottom) full-field displacement and strain field data for the $10 \times 10$ tessellation of unit cell geometry \#2 subjected to displacement-controlled uniaxial tension test.} \label{fig: full field comparison geom02}
	\end{center}
\end{figure}

\begin{figure}[!htb]
	\begin{center}
		\centering 
		\includegraphics[width =0.99\textwidth]{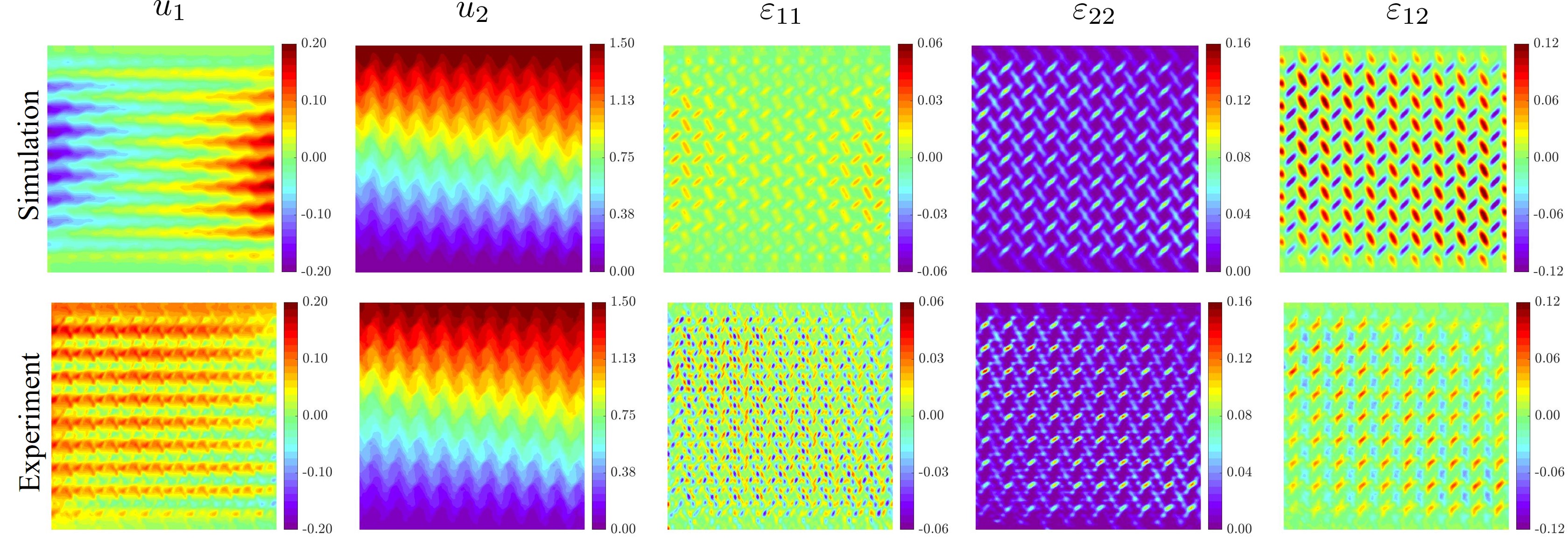}
	\caption{Comparison between numerical (top) and experimentally measured (bottom) full-field displacement and strain field data for the $10 \times 10$ tessellation of unit cell geometry \#3 subjected to displacement-controlled uniaxial tension test.} \label{fig: full field comparison geom03}
	\end{center}
\end{figure}

\begin{figure}[!htb]
	\begin{center}
		\centering 
		\includegraphics[width =0.99\textwidth]{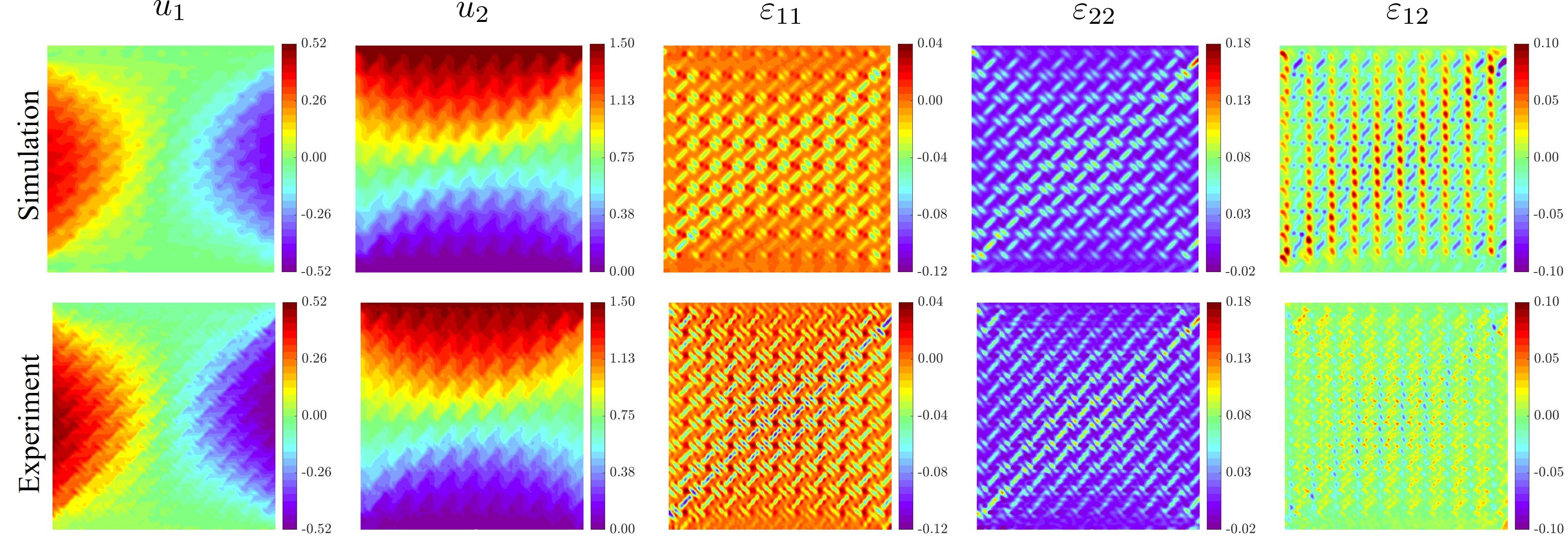}
		\caption{Comparison between numerical (top) and experimentally measured (bottom) full-field displacement and strain field data for the $10 \times 10$ tessellation of unit cell geometry \#4 subjected to displacement-controlled uniaxial tension test.} \label{fig: full field comparison geom04}
	\end{center}
\end{figure}

\newpage
\section{Comparison of simulated and experimentally measured displacement fields after postprocessing}

\begin{figure}[!htb]
	\begin{center}
		\centering 
        \includegraphics[width=0.65\textwidth]{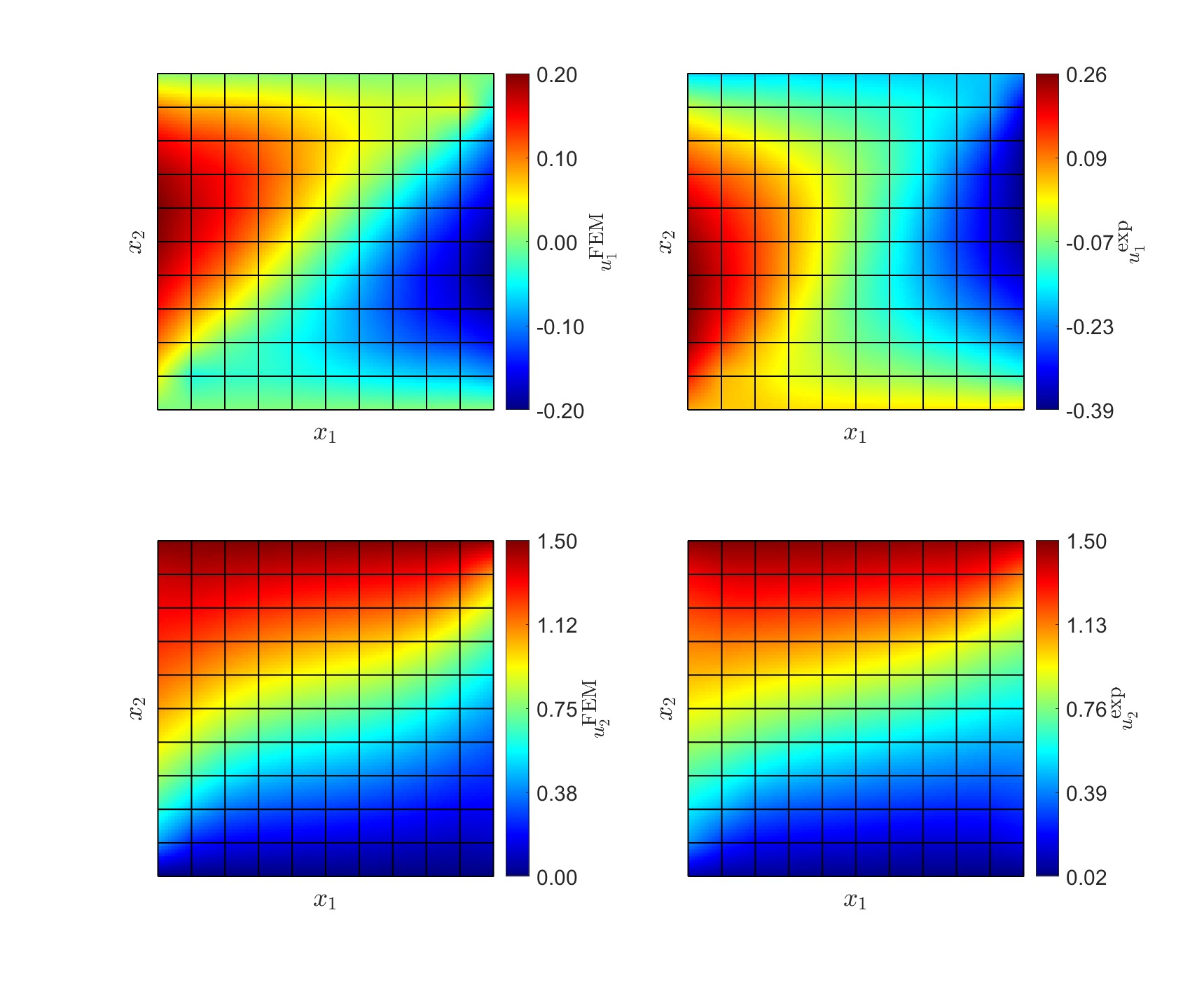}
		\caption{Comparison between numerical (top) and experimentally measured (bottom) full-field displacement and strain field data for the $10 \times 10$ tessellation of unit cell geometry \#1 subjected to displacement-controlled uniaxial tension.}
		\label{fig:displacement_FEMvsEXP_geom01}
	\end{center}
\end{figure}

\begin{figure}[!htb]
	\begin{center}
		\centering 
        \includegraphics[width=0.65\textwidth]{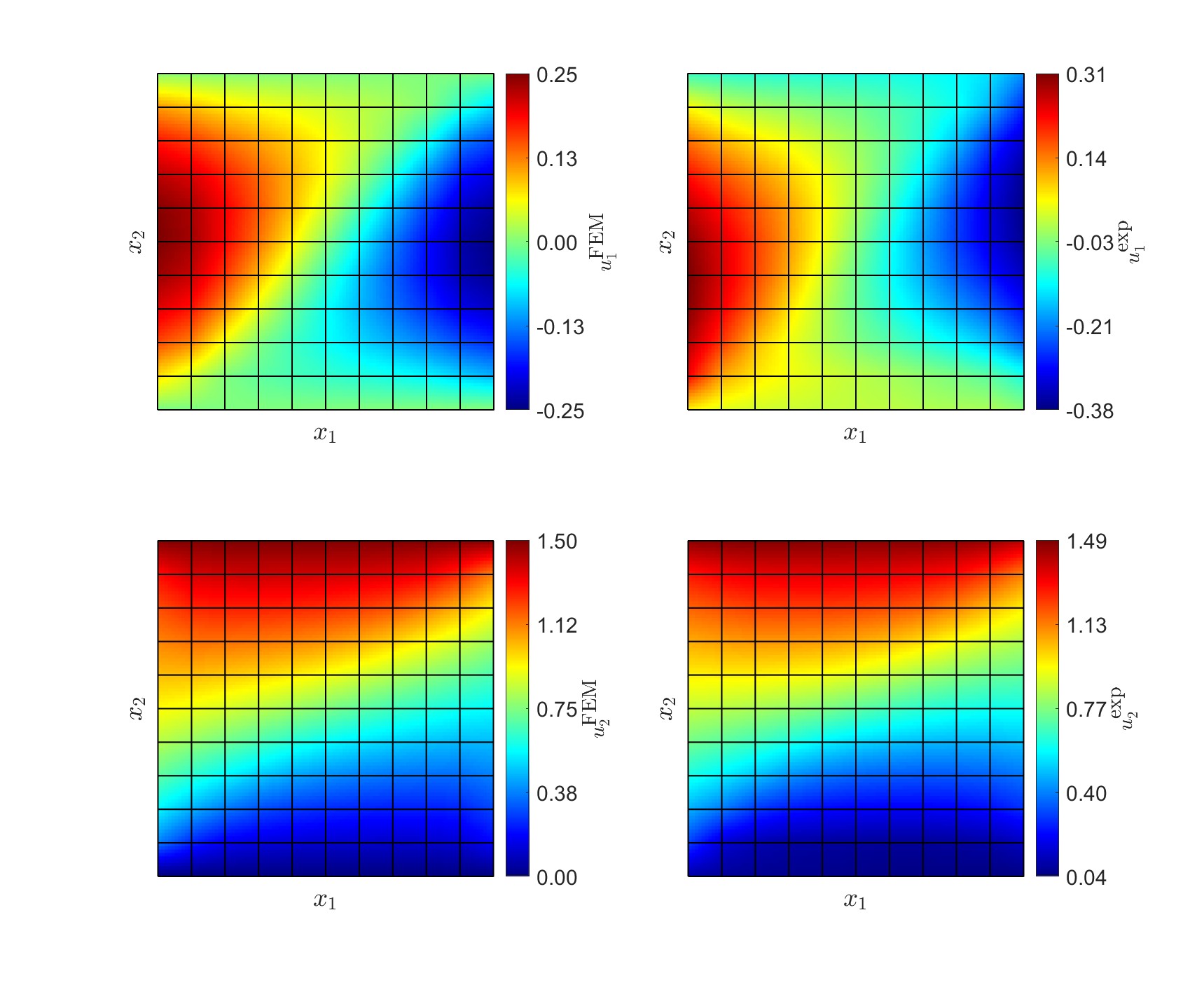}
		\caption{Comparison between the synthetic (left) and experimentally measured (right) displacement fields of the heterogeneous structure made of geometry \#2. Note that bilinear polynomials are used to interpolate the displacement data at the unit cell corners.}
		\label{fig:displacement_FEMvsEXP_geom02}
	\end{center}
\end{figure}

\begin{figure}[!htb]
	\begin{center}
		\centering 
        \includegraphics[width=0.65\textwidth]{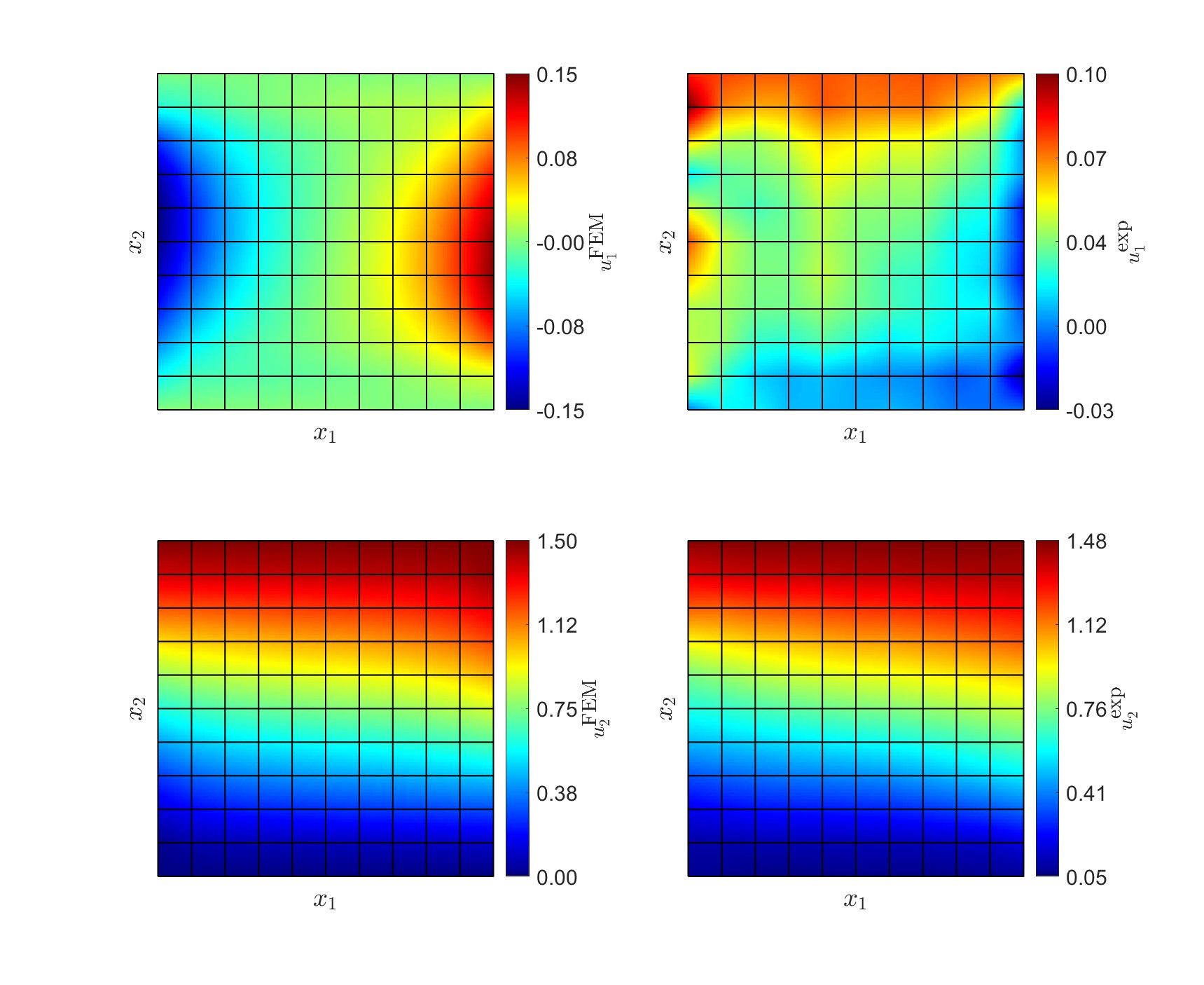}
		\caption{Comparison between the synthetic (left) and experimentally measured (right) displacement fields of the heterogeneous structure made of geometry 3. Note that bilinear polynomials are used to interpolate the displacement data at the unit cell corners.}
		\label{fig:displacement_FEMvsEXP_geom03}
	\end{center}
\end{figure}

\begin{figure}[!htb]
	\begin{center}
		\centering 
        \includegraphics[width=0.65\textwidth]{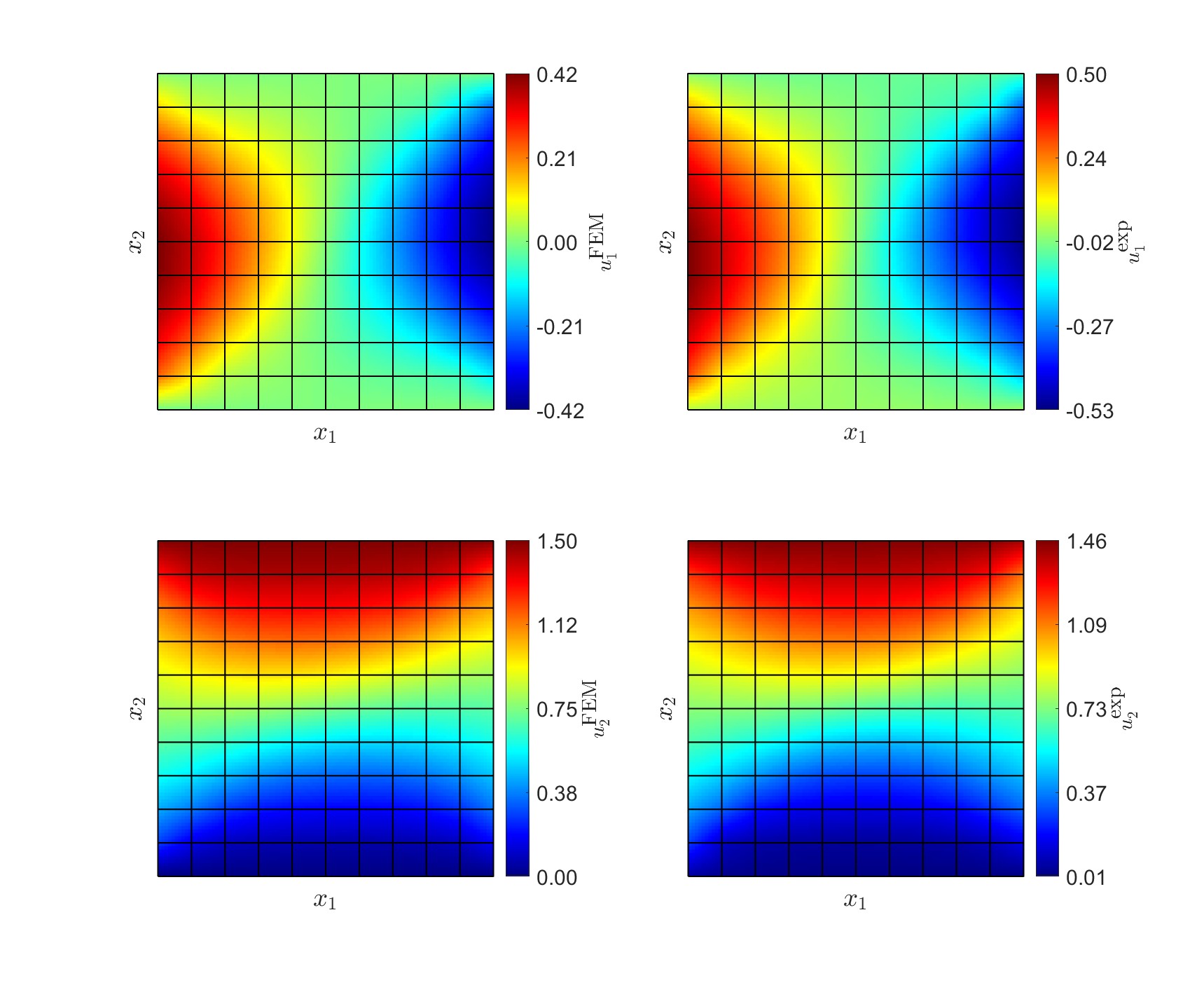}
		\caption{Comparison between the synthetic (left) and experimentally measured (right) displacement fields of the heterogeneous structure made of geometry 4. Note that bilinear polynomials are used to interpolate the displacement data at the unit cell corners.}
		\label{fig:displacement_FEMvsEXP_geom04}
	\end{center}
\end{figure}

\end{document}